\documentclass{aa}
\usepackage{graphicx}
\usepackage{amsmath}
\usepackage{amssymb}
\usepackage{bm}
\usepackage{natbib}
\usepackage{txfonts}
\begin{document}
\title{On the relevance of Subcritical Hydrodynamic Turbulence to Accretion Disk Transport}

%\subtitle{Resolution Effects in Cyclonic and Anticyclonic
%Subcritical Shear Flows}

   \author{G.\ Lesur\inst{1}
          \and
          P-Y.\ Longaretti\inst{1}
          }

   \offprints{G. Lesur}

   \institute{Laboratoire d'Astrophysique Observatoire de Grenoble
              BP 53 38041 Grenoble CEDEX 9\\
   \email{geoffroy.lesur@obs.ujf-grenoble.fr \& pierre-yves.longaretti@obs.ujf-grenoble.fr}
          }

   \date{Received June 23, 2005 / Accepted September 13, 2005}

   \abstract{Hydrodynamic unstratified keplerian flows are known
to be linearly stable at all Reynolds numbers, but may
nevertheless become turbulent through nonlinear mechanisms.
However, in the last ten years, conflicting points of view have
appeared on this issue. We have revisited the problem through
numerical simulations in the shearing sheet limit. It turns out
that the effect of the Coriolis force in stabilizing the flow
depends on whether the flow is cyclonic (cooperating shear and
rotation vorticities) or anticyclonic (competing  shear and
rotation vorticities); keplerian flows are anticyclonic. We have
obtained the following results:

i/ The Coriolis force does not quench turbulence in subcritical
flows; however, turbulence is more efficient, and much more easily
found, in cyclonic flows than in anticyclonic ones.

ii/ The Reynolds number/rotation/resolution relation has been
quantified in this problem. In particular we find that the
resolution demand, when moving away from the marginal stability
boundary, is much more severe for anticyclonic flows than for
cyclonic ones. Presently available computer resources do not allow
numerical codes to reach the keplerian regime.

iii/ The efficiency of turbulent transport is directly correlated
to the Reynolds number of transition to turbulence $Rg$,
in such a way that the Shakura-Sunyaev parameter $\alpha\sim
1/Rg$. This correlation is nearly independent of the flow
cyclonicity. The correlation is expected on the basis of generic
physical arguments.

iv/ Even the most optimistic extrapolations of our numerical data
show that subcritical turbulent transport would be too inefficient
in keplerian flows by several orders of magnitude for
astrophysical purposes. Vertical boundary conditions may play
a role in this issue although no significant effect was found in
our preliminary tests.

v/ Our results suggest that the data obtained for keplerian-like
flows in a Taylor-Couette settings are largely affected by
secondary flows, such as Ekman circulation.

   }

   \titlerunning{Subcritical Turbulence in Rotating Shear Flows}
   \authorrunning{Lesur and Longaretti}

   \maketitle

\section{Introduction}\label{intro}

The question of the existence and physical origin of turbulence in
accretion disks has been lively debated for a number of decades.
Generally speaking, there are a priori two basic ways in which an
accretion disk can become turbulent. In the first way, some linear
instability is present in the flow, and its nonlinear development
eventually drives turbulence. In the second one, the flow is
linearly stable, and undergoes a direct laminar-turbulent
transition once a certain threshold in Reynolds number is reached.
The first type of transition to turbulence is called
supercritical, and the second, (globally) subcritical.

Global instabilities (such as the \citealt{PP84} instability) seem
unpromising to drive turbulence \citep{B87,H91}. As for local
instabilities, an astrophysically important example of
supercritical transition is provided by the magneto-rotational
instability (MRI) which has been extensively studied following the
pioneering work of Balbus, Hawley and their collaborators
(\citealt{BH91}; \citealt{HGB95}; see \citealt{B03} for a recent
review). The turbulent transport induced by this instability is by
now characterized in a number of instances, and has been called
upon even when only some fraction of the disk is ionized, as in
the midplane region of YSOs inner disks --- the dead-zone
(\citealt{G96}; \citealt{FS03}). However, the reduced efficiency
of the transport in this case, as well as the possible existence
of disks which may not support MHD phenomena at all, has prompted
some upsurge of interest in purely hydrodynamic instabilities. A
local, baroclinic-like instability has been observed in global
simulations by \cite{KB03}. Local stability analyzes
\citep{K04,JG05a} find transient instability in this context, but
shearing box simulations indicate that this does not drive
turbulence \citep{JG05b}. \cite{U03} discusses an instability
related to vertical shear and heat transport of the
Goldreich-Schubert type \citep{GS67}; however, this instability
produces only a rather weak radial transport \citep{AU04}. More
recently, \citet{DMNRHZ05} and \citet{SR05} have discussed
an instability arising when both the fluid differential rotation
and vertical stratification are stabilizing according to the H\o
iland criterion. However, it seems that this instability is
connected to the presence of walls, and is dynamically important
only when the inter-wall distance is small enough for a
resonant-like interaction to take place\footnote{We thank
St\'ephane Le Dizes for bringing this point to our attention.}
\citep{S81}, otherwise disturbances are confined to the near
boundary zone; a related result has recently been found in
the astrophysics literature \citep{U05}. Earlier analytic and
numerical investigations have shown this instability to be absent
in local disk models \citep{GB01,BD01,RAS02}. Note finally that
vertical convection in a stratified disk can in principle also
drive turbulence; however, it induces inwards transport instead of
the required outwards one (\citealt{C96}; \citealt{SB96}).
Therefore, no local instability has yet been found in the
hydrodynamic regime, which would explain the turbulent transport
taking place in accretion disks.

Subcritical transition to turbulence is the subject of the present
work. The non-rotating plane Couette flow provides a classical
(and to date the best understood) example of a system undergoing a
subcritical transition. Although the nature and mechanism of the
transition remained elusive for decades, it has been identified in
the recent years, in laboratory experiments (\citealt{DHB92};
\citealt{DD95a}; \citealt{DD95b}; \citealt{BDD97}), numerical
simulations (\citealt{HKW95}; see also \citealt{SE97} and
\citealt{EM99}), and theoretical analyzes (in particular
\citealt{W97}; \citealt{W03}). Earlier investigations of
the problem have focused on the role of nonlinear instabilities in
subcritical shear flows, based on Landau-like toy-models on the
one hand (e.g., \citealt{DR81} and references therein), and
analysis of the linear stability of finite amplitude defects in
the flow profile on the other (\citealt{LK88}, \citealt{DZ91},
\citealt{D93}); unfortunately, such analyzes yield little
information on the existence and location of the turbulent state
in parameter space and on the turbulent transport efficiency,
unless further ad hoc assumptions are made.

In any case, on the basis of the empirically observed subcritical
transition in laboratory flows, it was suggested that a similar
process is relevant in accretion disks (\citealt{SSL78}), in spite
of their very different prevailing physical conditions. This
suggestion was tested and challenged in a series of numerical
simulations performed by \cite{BHS96} and \cite{HBW99}, in the
shearing sheet limit. Transition to turbulence was not found in
these simulations for keplerian-like flows. The simulations were
performed with two different finite difference codes (a PPM type
code, and the ZEUS code), up to a resolution of $256^3$. These two
works concluded that a stabilizing Coriolis force prevents the
existence of turbulence in the simulated flows, except in the
immediate vicinity of the linear marginal stability limits.

This conclusion was in turn questioned by \cite{RZ99}, on the
basis of the Taylor-Couette experiments performed by \cite{W33}
and \cite{T36}. These experimental results display a subcritical
transition to turbulence in presence of a stabilizing Coriolis
force. Also, new sets of experiments have been carried out in
order to bring the experimental conditions closer to the ones
prevailing in a keplerian flow. Namely, a Taylor-Couette apparatus
was used in conditions of radially decreasing angular velocity and
radially increasing specific angular momentum. Turbulence was
again found for high enough Reynolds numbers (\citealt{R01};
\citealt{RDDZ01}) but the results are not unambiguous, as the
potential role of secondary flows induced by the boundary
conditions in the experiments, such as Ekmann's circulation, is
unclear, in spite of the attention devoted to this point in the experiments.
In any case, a subcritical transition is also found in
all experiments of shear flows on which a linearly stabilizing
Coriolis force is superimposed (\citealt{LD05}).

\cite{L02} has argued from a phenomenological analysis that the
lack of turbulence in the simulations performed to date was due to
a lack of resolution, as the Coriolis force may increase the range
of scales that need to be resolved for a subcritical turbulent
transition to show up. On the other hand, on the basis of a newly
developed Reynolds stress closure scheme (\citealt{O03}),
\cite{GO05} find that keplerian flows may or may not be turbulent
depending on the parameters of the scheme. For their favored
choice of parameters, unbounded keplerian flows are not turbulent,
on the contrary to linearly stable, wall-bounded Taylor-Couette
flows.

The recent astrophysical literature on the problem of subcritical
transition has also focused on the concept of transient growth in
keplerian flows (\citealt{CZTL03}; \citealt{TCZCL03};
\citealt{Y04}; \citealt{UR05}; \citealt{MAN05}; \citealt{AMN05}).
Due to the nonnormal character of the Navier-Stokes equation,
linear modes can transiently be strongly amplified in shear flows,
although on the long run they must viscously decay. It has been
argued that this transient growth can be relevant to astrophysical
disks in two different ways. First, 3D turbulence (or an external
forcing) can couple to large scale 2D structures; the
(statistical) amplitude of these structures can be large, under
the combined action of this coupling, of transient growth and of
viscous decay, and these 2D structures may contribute to the
overall transport in the disk (\citealt{IK01}). Secondly, a large
transient growth has been invoked in the bypass scenario of
transition to turbulence, which involves an interplay between
nonnormality and nonlinearity (see, e.g., \citealt{G00};
\citealt{BG99}). \cite{W95} has emphasized the key role played by
nonlinear interactions in the context of the recently identified
turbulent self-sustaining process of non-rotating plane Couette
flows (\citealt{HKW95}; \citealt{W97}). Even though transient
growth explains the strong modulations of the streamwise velocity
from relatively weak streamwise rolls involved in this
self-sustaining mechanism, the existence and properties of the
turbulent basin of attraction for the full nonlinear dynamics
are apparently poorly constrained by the nonnormal linear
problem characteristics.

Our present understanding of the possible existence of a
dynamically significant subcritical turbulent transition in
accretion disks is unsatisfying in several respects, calling for a
reinvestigation of the problem. On the one hand, the relevance of
the available laboratory experiments to accretion disk turbulence
is at best unclear, as will be shown in the course of the
present work (for a different opinion, see \citealt{HDH05}). On
the other hand, the absence of subcritical turbulence in the
shearing sheet local model of accretion disks used by \cite{BHS96}
and \cite{HBW99} may be an effect of various numerical
limitations, namely, algorithm choice, limited resolution, nature
of the boundary conditions, imposed aspect ratio and initial
conditions of the simulations. Of these options, only the first
two have been partially addressed in these previous
investigations, leading to questions concerning the ``effective
Reynolds number" of the performed simulations --- an ill-defined
process-dependent concept, that we shall clarify in the context of
the present problem. Following the suggestion of \cite{L02}, the
primary aim of the present work is to investigate in a more
systematic way, through numerical simulations of plane parallel,
rotating shear flows, the effects of finite resolution on the
results. The effects of the other factors listed above are also
somewhat explored, but to a lesser extent. Both cyclonic and
anticyclonic rotation are considered; although cyclonic rotation
is not relevant to accretion disks, it turns out that cyclonic
flows behave very differently from anticyclonic ones, opening some
interesting perspective into the nature of the problem.

This paper is organized as follows. Section \ref{equ} collects the
background material relevant to the problem. First, the form of
the equations solved is provided, and the global energy budget
recalled, before discussing linear stability limits. The section
is concluded by a summary of the effect of a stabilizing rotation
in shear flows as characterized by the available laboratory
experiments. The next section presents the various codes used in
this work, and the numerical results obtained with them. Section
\ref{discussion} discusses various aspects of our numerical
results, most notably the role of resolution and boundary
conditions on the numerical side, the role of the Coriolis force,
the underlying phenomenological picture, and the astrophysical
implications, on the physical side. A summary is provided in
section \ref{conclusion}, along with an outlook on the question of
turbulence in accretion disks.

\section{Rotating plane shear flows: a summary}\label{RPSF}

The present investigation is concerned with the nonlinear
instability of laminar flows characterized by a uniform shear, in
the presence of a uniform global rotation. The direction of the
flow is identified with the $x$ axis (streamwise direction), and
the direction of the shear with the $y$ axis (shearwise
direction); rotation is applied along the $z$ axis (spanwise
direction). The laminar flow ${\mathrm u}_L$ is invariant in the
streamwise and spanwise directions (in particular, the vertical
stratification expected in a real disk is ignored): ${\mathrm
u}_L=U(y){\mathrm e}_x$.

\begin{figure}[htb]
\centering
\includegraphics[scale=0.4]{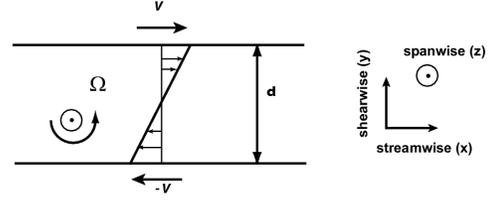}
\caption{\small Sketch of the configuration of rotating plane
shear flows.}\label{RPC}
\end{figure}

Such a flow can be used to numerically model either a local
portion of an accretion disk, or experiments on rotating plane
Couette flows, depending on the nature of the applied boundary
condition in the shearwise direction (in practice, either rigid or
shearing sheet; see next section). The configuration is
represented on Fig.~\ref{RPC}.

\subsection{Equations of motion}\label{equ}

The most useful form of the Navier-Stokes equation, for our
present purpose, is obtained by separating the laminar flow
${\mathbf u}_L$ and the deviation from laminar ${\mathbf w}$ in
the total velocity ${\mathbf u}$ in the rotating frame, leading to

\begin{eqnarray}\label{split-RPC}
  \frac{\partial{\mathbf w}}{\partial t}+
  {\mathbf w}\cdot{\bm\nabla}{\mathbf w} =
  S\cdot y \frac{\partial{\mathbf w}}{\partial x}
  & + & (2\Omega+S) w_y {\mathbf e}_x
  - 2\Omega w_x {\mathbf e}_y \nonumber \\
   & - & \frac{{\bm\nabla}\delta\pi}{\rho}+\nu\Delta{\mathbf w},
\end{eqnarray}

\noindent where the gradient terms balancing the laminar flow
Coriolis force has been subtracted out to form the effective
generalized pressure $\delta\pi$ (which therefore absorbs the
equilibrium centrifugal, gravitational and/or pressure force term,
depending on the considered equilibrium problem); $\Omega$ is the
flow rotation velocity in an inertial frame, and $S=-dU/dy$ is the
shear. The convention adopted here is that the sign of $S$ is
chosen to be positive when the flow is cyclonic, i.e., when the
contributions of shear and rotation to the flow vorticity have the
same sign. With our choice of axes, this implies that
$S=-2S_{xy}$, where $S_{ij}=1/2(\partial_i u_{L,j}+\partial_j
u_{L,i})$ is the usual deformation tensor. The system is closed
either with the usual continuity equation supplemented by a
polytropic equation of state, or, for simplicity, through an
incompressibility assumption (${\bm\nabla}\cdot{\mathbf w}=0$).

The relevant global time-scales of the problem are the shear
time-scale $t_s=|S^{-1}|$, the viscous one $t_\nu=d^2/\nu$ ($d$ is
the gap in the experiment, or the shearwise size of the shearing
sheet box), and the rotation time-scale related to the Coriolis
force $t_\Omega=(2\Omega)^{-1}$; they relate to the advection
term, the viscous term, and the Coriolis force term, respectively.
Correlatively, the flow is described by two dimensionless numbers,
the Reynolds number

\begin{equation}\label{reyn}
Re=t_\nu/t_s=|S|d^2/\nu,
\end{equation}

\noindent and the rotation number

\begin{equation}\label{rot}
R_\Omega= \mathrm{sgn}(S) t_s/t_\Omega=2\Omega/S.
\end{equation}

\noindent For Keplerian flows, $R_{\Omega}=-4/3$. More generally,
if one assumes that the large scale rotation of an astrophysical
disk follows a power-law, $\Omega(r)\propto r^{-q}$, one locally
has $R_\Omega=-2/q$ in the disk.

Note that our Reynolds number is defined on the outer
scales, and not on the turbulent ones, such as, e.g., the Taylor
microscale. Large values ($\sim 10^4$) of this number are involved
in the problem investigated here; the correlative numerical
requirements are discussed in section \ref{eff-re}.

\subsection{Energy budget}\label{energy}

As the global energy budget plays some role in the discussion of
the results, it is rederived here. In the following equations, the
bracket notation refers to a volume average of the bracketed
quantity. The averaging volume is the simulation one, and
shearing-sheet boundary conditions are assumed in the derivation,
for definiteness. For the kinetic energy in the streamwise and
shearwise directions, one finds:

\begin{align}\label{wx2av}
  \frac{\partial}{\partial t}\Big\langle\frac{
  w_x^2}{2}\Big\rangle = &
  S\left(R_\Omega+1\right)\langle w_x w_y\rangle
 \nonumber\\ & -\Big\langle\frac{w_y}{\rho}
  \frac{\partial\delta \pi}{\partial x}\Big\rangle + \nu\langle w_x\Delta {
  w}_x\rangle,
\end{align}

\begin{align}\label{wy2av}
  \frac{\partial}{\partial t} \Big\langle\frac{w_y^2}{2}\Big\rangle
  = & -SR_\Omega\langle\ w_x w_y\rangle \nonumber\\ &
  -\Big\langle\frac{ w_y}{\rho}\frac{\partial\delta
  \pi}{\partial y}\Big\rangle + \nu\langle w_y\Delta w_y\rangle.
\end{align}

\noindent Instead of the vertical equation, it is more instructive
to write down the total kinetic energy equation:

\begin{equation}\label{w2av}
  \frac{\partial}{\partial t}\Big\langle\frac{
  {\bm w}^2}{2}\Big\rangle =
  S\langle w_x w_y\rangle - \epsilon,
\end{equation}

\noindent where

\begin{equation}\label{epsilon}
  \epsilon=\nu\sum_i\langle({\bm\nabla} w_i)^2\rangle
\end{equation}

\noindent is the usual energy injection rate of turbulence cascade
arguments\footnote{Because the rate of energy transfer in scale is
constant in a Kolmogorov-like argument, the injection rate is
directly related to the small-scale dissipation rate.}. In this
last equation the incompressibility condition and the boundary
conditions have been used in the reexpression of the pressure
term, and an integration by part has been performed on the viscous
term (a constant kinematic viscosity $\nu$ is assumed).

In statistical steady-state, Eq.~(\ref{w2av}) reduces to,

\begin{equation}\label{w2avb}
  S\langle w_x
  w_y\rangle = \epsilon.
\end{equation}

As pointed out by \cite{BHS96}, the fact that $\epsilon > 0$
implies that in steady state, the shear rate and the Reynolds
stress responsible for radial transport have identical signs. This
result has a direct physical interpretation: the imposed shear
prevents the flow to be in global thermodynamic equilibrium.
Nevertheless, the flow tries to restore this global
equilibrium by radially transporting momentum through the
turbulent Reynolds stress from regions of larger momentum to
regions of lower momentum, consistently with Eq.~(\ref{w2avb}).

Note finally that, in Eqs.~(\ref{wx2av}) and (\ref{wy2av}), the
pressure-velocity correlation terms \textit{cannot} be neglected,
as they are of the order of the cascade energy injection term
$\epsilon$. This is almost unavoidable, as pressure is the only
force that can provide for the acceleration of fluid particles in
turbulent motions. As a matter of fact, the energy budget of any
particular velocity component depends critically on the behavior
of the velocity-pressure correlations, which are notoriously
difficult to model \citep{S91}. Ignoring this term in the analysis
of the energetics therefore leads to dubious or erroneous
conclusions.

\subsection{Linear stability limits}\label{stablin}

Surprisingly enough, the question of the linear stability limits
of the simple rotating shear flows considered here is not
completely solved to date. Focusing for the time being on purely
streamwise-independent perturbations, instability with respect to
local perturbations follows when (\citealt{P69}; \citealt{LC97};
\citealt{SJ00})

\begin{equation}\label{stab}
R_\Omega(R_\Omega+1) < 0,
\end{equation}

\noindent or, equivalently, $-1 < R_\Omega < 0$.

In plane Couette flows, it has been proven that $R_\Omega^+\equiv
0$ is the correct cyclonic marginal stability limit for non
streamwise-invariant perturbations as well, at all Reynolds
numbers (\citealt{R73}). No such generic proof exists at the
anticyclonic marginal stability limit ($R_\Omega^-\equiv -1$).
However, various linear and nonlinear numerical investigations
suggest that this is indeed the case (\citealt{CBSJ94};
\citealt{KLJ96}; \citealt{BA97}). These results belong to plane
Couette flows with rigid boundary conditions in the shearwise
direction, but tend to prove that a local criterion captures the
correct stability limit, as observed, e.g., in the simulations of
\cite{BHS96} and \cite{HBW99}.

The physics behind Eq.~(\ref{stab}) can be captured by a displaced
particle argument (\citealt{TD81}; \citealt{T92}). This argument
is reproduced in Appendix \ref{app:disp} for the reader's
convenience. Note that Eq.~(\ref{stab}) is identical to Rayleigh's
specific angular momentum criterion for the centrifugal
instability, as the usual epicyclic frequency reads $\kappa^2=S^2
R_\Omega(1+R_\Omega)$. However, in the plane shear flow limit of
cylindrical flows, the concept of specific angular momentum used
in the derivation of Rayleigh's criterion no longer has meaning,
so that one must follow a different route, as done here. Note also
that, consequently, the Rayleigh criterion for the centrifugal
instability in the inertial frame can also be understood from the
action of the Coriolis force in the rotating frame (a somewhat
surprising, although not new conclusion), as the displaced
particle argument of Appendix \ref{app:disp} is readily extended
to cylindrical flows.

\subsection{Subcritical transition in rotating plane Couette
flows: a summary of relevant experimental results}\label{data}

In the laboratory, non-rotating plane Couette flows undergo a
subcritical transition to turbulence at $Re\simeq 1500$. The
transition Reynolds number steeply increases if a stabilizing
rotation and/or a curvature is superimposed on the flow. The
conceptually cleanest way to add rotation to a plane Couette flow
is to place a plane Couette apparatus on a rotating table. Also,
by considering a Taylor-Couette apparatus with varying gap width
and independently rotating cylinders, one obtains a flow in which
both rotation and curvature effects can be studied, and which
reduces to a rotating plane Couette flow in the narrow gap limit.
For a more complete discussion of the distinction and
characterization of rotation and curvature in Taylor-Couette
experiments, and of the related experimental data, the reader is
referred to \cite{LD05}.

\begin{figure}[htb]
\centering
\includegraphics[scale=0.5]{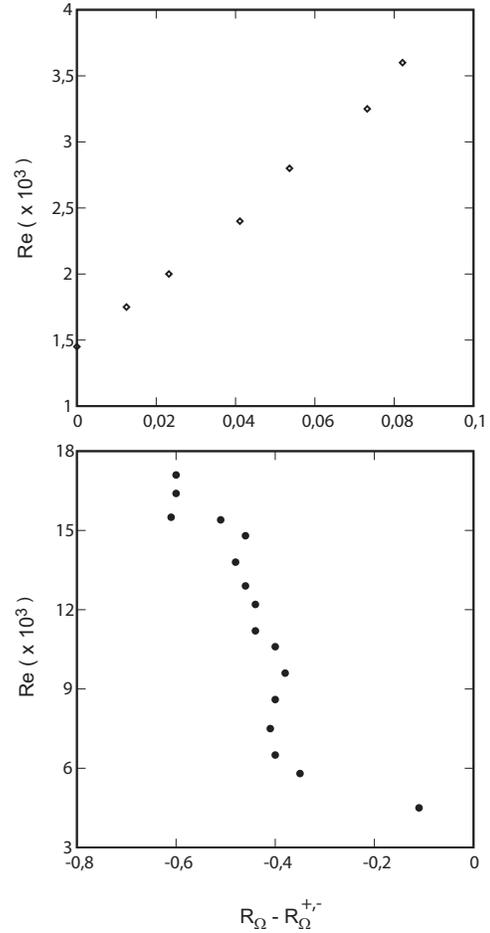}
\caption{\small Data on the Reynolds number of subcritical
transition to turbulence as a function of the rotation number
$R_\Omega$, measured from the appropriate marginal stability limit
$R_\Omega^\pm$ (see text). Top panel: cyclonic plane Couette flow
(data from \citealt{TA96}). Bottom panel: anticyclonic
Taylor-Couette flow (data from \citealt{R01}). The anticyclonic
data are more difficult to collect, and consequently
noisier.}\label{rotdata}
\end{figure}

For the range of parameters studied to date in the experiments, it
turns out that rotation and curvature effects on the transition
Reynolds number are superposed in an mostly additive way, so that
both plane Couette flows and Taylor-Couette flows can in principle
be used to characterize the effect of rotation. Concerning
cyclonic flows, the only directly relevant data have been
collected by \cite{TA96} with the help of a plane Couette flow
apparatus placed on a rotating table. For anticyclonic flows, the
only available experiments are those of Richard and coworkers
(\citealt{R01}; \citealt{RDDZ01}), who used a Taylor-Couette
apparatus. The range of rotation number $R_\Omega$ explored in
these experiments is $0$
--- $0.1$ for cyclonic rotation, and $-1.6$ --- $-1$ for
anticyclonic rotation. The data are shown on Fig.~\ref{rotdata}

The important point to note here is the steep dependence of the
transition Reynolds number with the ``distance" to marginal
stability, with a typical slope $|\Delta R_g|/|\Delta
R_\Omega|\sim 10^4$ --- $10^5$.

\section{Numerical codes, strategy, and results}\label{numerics}

In the present work, we are concerned with rotating, unstratified
uniform shear flows. Periodic boundary conditions hold in the
direction of the flow ($x$ axis) and the ``vertical" direction
($z$ axis), and either shearing sheet or rigid boundary conditions
are applied in the direction of the shear ($y$ direction). The
vertical axis is also the axis of rotation of the flow. The
shearing sheet boundary conditions are described in detail by
\cite{HGB95}. Shearing sheet flows thus modelled can be viewed as
a local approximation of disk flows, while the use of mixed
rigid-periodic boundary conditions is appropriate to numerically
represent the rotating plane Couette flows of laboratory
experiments, as routinely done in the fluid mechanics community.

\subsection{Numerical codes}\label{codes}

Two different 3D codes have been written for the present work: a
finite difference compressible code, similar to ZEUS
(\citealt{SN92}), but restricted to the cartesian geometry, and
rigid-periodic or shearing sheet boundary conditions; and a 3D
incompressible Fourier code, in cartesian geometry, and
implementing only the shearing sheet boundary conditions.
An explicit kinematic viscosity term is added in both
codes, upon which the Reynolds number is defined. Both codes were
parallelized using the Message Passing Interface.

The shearing sheet boundary conditions induce some changes with
respect to a standard Fourier code. As a matter of fact, while we
were developing this code, the work by \cite{UR05} appeared, which
implements the same technique. Therefore, our description of the
required changes will be brief, and we refer the reader to this
recent paper for details.

To get effective periodic boundary conditions on the 3 axes, one
needs to write Eq.~(\ref{split-RPC}) in the sheared frame defined
by:

 \begin{eqnarray}\label{shearaxes}
 t'&=&t\\
 x'&=&x+S\cdot y\cdot t\\
 y'&=&y\\
 z'&=&z.
 \end{eqnarray}

\noindent In this shearing frame, Eq.~(\ref{split-RPC})
(supplemented by the incompressibility condition) becomes:

 \begin{eqnarray}
 \label{hydro3}
 \frac{\partial \bm{w}}{\partial t'}+\bm{w}\cdot\tilde{\bm\nabla}
 \bm{w}&=&-\frac{\tilde{\bm\nabla} \delta\pi}{\rho} -2\Omega\, w_x \bm{e_y}+(2\Omega +S)w_y
 \bm{e_x}\nonumber\\ & & + \nu \tilde{\Delta} \bm{w}\\
 \tilde{\bm \nabla} \cdot \bm{w} & = & 0.
 \end{eqnarray}

\noindent in which $\tilde{\bm \nabla}=\partial_{x'}
\bm{e_{x'}}+(\partial_{y'}-St'\partial_{x'})\bm{e_{y'}}+\partial_{z'}\bm{e_{z'}}$
and $\tilde{\Delta}=\tilde{\bm \nabla}\cdot \tilde{\bm \nabla}$.

Since the shearing box is a periodic box in the shearing frame,
this last formulation of the Navier-Stokes equation can be written
in 3D-Fourier Space. Defining

 \begin{equation}\label{shear-k}
 \bm{\mu}=\bm{k}-Stk_x\bm{e_y},
 \end{equation}

\noindent one finally obtains:

 \begin{eqnarray}\label{NS-Fourier}
 \frac{\partial \hat{\bm{w}}}{\partial t'}+i\bm{\mu}\cdot \widehat{\bm{w}\otimes
 \bm{w}}&=&-i\bm{\mu} \frac{\widehat{\delta\pi}}{\rho} -2\Omega\,
 \hat{w}_x \bm{e_y}+(2\Omega + S)\hat{w}_y \bm{e_x}\nonumber\\
 & &-\nu \mu^2 \bm{w}\\
 \bm{\mu} \cdot \hat{\bm{w}}&=&0
 \end{eqnarray}

These are the equations actually used in our spectral code. The
nonlinear term is computed using the 2/3 dealiasing rule with a
pseudo-spectral method (see e.g \citealt{P02} for a description of
this point) and each time-step is evaluated using a
$4^{\mathrm{th}}$ order Runge Kutta Scheme. One should note that a
k-wave in the sheared frame actually appears as a
$\mu(t)$-wave in the steady frame. Then, as time goes on in the
simulation, the k-grid describes higher spatial frequency in the
steady frame and consequently, the large scales are not computed
anymore. Since nonlinear coupling limits the shearing of any
wave-number, a remap procedure is periodically applied all along
the simulation, and prevents to loose information on the large
scale\footnote{We thank Achim Wirth for pointing out this
reference to us.} (\citealt{R81}). This kind of algorithm has been
extensively described by \cite{UR05} using a 2D spectral code and
the reader should refer to this publication for technical details
on the remap procedure.

The choice of these two codes was made first for purposes of
comparison with previous work, and secondly to allow us to
cross-check the potential limitations of one code against the
other; e.g., the shearing sheet boundary conditions and sheared
spatial basis Eq.~(\ref{shear-k}) have their own limitations, as
the sheared basis forms a complete basis for shearing sheet
boundary conditions, but only for these conditions.

The three codes were tested in a variety of ways. The first test
was to reproduce the non-rotating plane Couette flow behavior
computed by \cite{HKW95}. This was done both with our finite
difference code, and with David Clarke's version of ZEUS3D, for
comparison purposes. We checked the non-linear transition
mechanism was well reproduced, with the corresponding Reynolds
number and aspect ratio, and that the two codes gave completely
consistent results. Then, the shearing sheet boundary conditions
were tested using these two finite difference codes and the
Fourier code. We have verified that mean turbulent quantities
(e.g., mean energy, mean transport, velocity maxima and minima)
and critical Reynolds number were statistically the same using the
different codes, for different rotation numbers, either cyclonic
or anticyclonic. This consistency holds over the $10^5-10^6$ time
steps of our simulations.

\subsection{Initial conditions and numerical strategy}\label{init}

The experimental results recalled in section \ref{data} suggest
that a steep dependence of the transition Reynolds number with the
rotation number may be the cause of the difficulty to find such a
transition in the previously published shearing sheet numerical
simulations. Accordingly, one of the major aims of this
investigation is to quantify the effect of the simulation
resolution on the determination of the transition Reynolds number
as a function of $R_\Omega$.

Now, one of the characteristic features of the subcritical
transition to turbulence is an observed spread in transition
Reynolds numbers, depending on the choice of initial conditions,
and a correlative large spread in turbulence life-times. This has
been documented both experimentally \citep{DM95} and numerically
\citep{FE05} in pipe flows, and guides to some extent our choice
of initial conditions and our numerical procedure. Indeed,
turbulent life-times typically vary from fast decay (survival for
less than one hundred dynamical times) to long or indefinite
survival (several thousands of dynamical times, with a clear
divergence at finite Reynolds number) over several orders of
magnitude of variation of the initial condition amplitude, but for
less than 50$\%
$ of variations of Reynolds number (see
\citealt{FE05}, Figs.~2 and 7).

It is reasonable to assume that this qualitative behavior is
generic. Consequently, we have chosen once and for all, fixed,
high amplitude initial conditions, to make our numerical runs more
directly comparable to one another upon variations of Reynolds
numbers. Furthermore, we consider that turbulence is long-lived if
it is not observed to decay for 100 or 200 shear times (depending
on the runs). This choice is a compromise between computational
time constraints, and accuracy in the determination of the
transition Reynolds number of indefinitely self-sustained
turbulence. In practice, simulations are performed in a cubic box
(the impact of this choice is discussed in the next section, to
some extent). The flow is adimensionalized with the only
dimensional quantities introduced in the problem: $S$ and $d$,
where $d$ is the simulation box size (or equivalently, by choosing
$|S|=1$ and $d=1$). The initial conditions used for all our
simulation are a random 3D excitation of the 10 largest Fourier
modes, with \textit{rms} fluctuations in velocity of order unity
in our chosen units. Other shapes of initial conditions were
tested such as white noise (all scales excited randomly) or
introducing large scale vortices in various directions with a
small superimposed noise. This produces no significant difference
once the flow is relaxed ($t \gtrsim 20 \,t_s$).

The numerical strategy adopted is then rather straightforward:
choosing a code, a resolution, a boundary condition (for the
finite difference code) and a Reynolds number, at fixed initial
conditions, the flow evolution is computed starting from the
marginal stability limit in rotation number $R_\Omega$ and
evolving the rotation number by (small) fixed steps every 100 or
200 shear times. According to the preceding discussion, this
allows us to reduce at maximum the number of runs and the run time
needed to observe systematic trends in the numerical results.

In this section, only shearing sheet boundary conditions are used.
We have also checked that the time required to dissipate the
turbulent energy of the flow assuming energy injection is stopped
(deduced from the $\epsilon$ term in Eq.~[\ref{w2av}]) is smaller
than $100 t_s$; this constraint is always satisfied by a large
margin in all our runs, implying that the deviations from laminar
motion that we observe are self-sustained (i.e., we do not observe
them because their dissipation time exceeds the run time).
Actually, once turbulence is lost in our simulations, the energy
in the velocity fluctuations always decreases rather fast, as can
be checked on Fig.~\ref{statcyc} for cyclonic flows. The same property
is found for anticyclonic flows, see section  \ref{res-anticyc}.

We conclude this section on our choice of the Mach number  ($Ma=L_y S/c_s$) for
our simulations with the compressible Zeus-like code. The type of
motions we are considering in these simulations reach at most a
small fraction of the boundaries relative velocity (normalized to
unity in this work). We found that a sound speed also normalized
to unity was a good compromise between limiting the effects of
compressibility (which eventually makes the turbulence
compressible and largely different in character when the Mach
number is too large), and the impact of the sound speed on the CFL
condition. Also, this value mimics the real role of
compressibility in a vertically stratified accretion disk.
Consequently, $Ma=1$ is imposed in all our compressible
simulations.

\subsection{Numerical results: cyclonic flows}\label{res-cyc}

\begin{figure}[htb]
  \centering
  \includegraphics[scale=0.45]{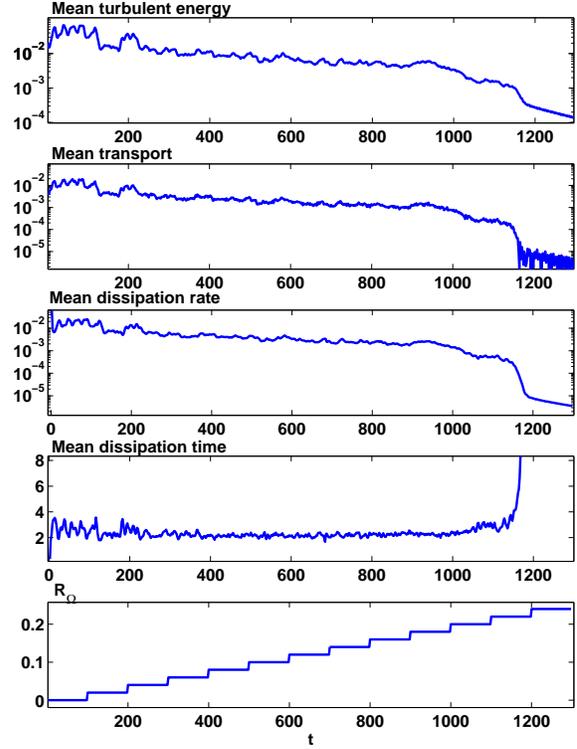}
  \caption{\small Example of the time evolution of a $64^3$
  (Re=12000) cyclonic flow run as computed by our Fourier code.
  The turbulent energy, transport and dissipation rate are the quantities
  involved in Eq.~(\ref{w2av}). The dissipation time follows from the turbulent energy
  and the dissipation rate. The bottom panel displays the evolution of the rotation
  number that is imposed in the course of the simulation.}\label{statcyc}
\end{figure}

On the cyclonic side, simulations are performed while maintaining
the rotation number $R_\Omega$ constant during 100 shear times;
then the rotation number is increased by steps of 0.02, starting
from the marginal cyclonic point $R_\Omega=0$. An example global
output of such a simulation is plotted on Fig.~\ref{statcyc} for
Re=12000. The relaminarization point is easily found since the
transition between the turbulent to laminar state is quite abrupt
(at $t=1150$ on Fig.~\ref{statcyc}). We define the last turbulent
point as the last rotation rate for which turbulence is sustained
for $100\,t_s$. For our example simulation, we find that the last
turbulent point at Re=12000 and $64^3$ resolution with our Fourier
code is $R_\Omega=0.2$.

\begin{figure}
  \centering
  \includegraphics[scale=0.35]{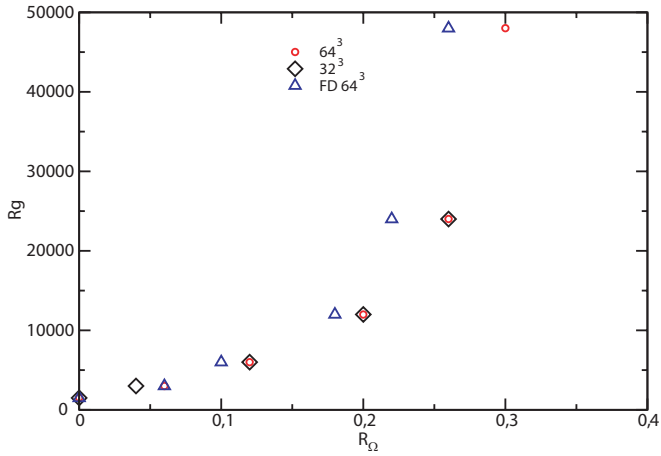}
  \caption{\small Transition Reynolds number $Rg$ as a function of the rotation
  number $R_\Omega$, with different resolutions and codes for
  shearing-sheet boundary conditions (cyclonic rotation). All points were
  obtained using our Fourier code except those labelled
  FD (finite difference) which use our ZEUS-like code.}\label{graphcyc}
\end{figure}

Using this kind of simulation, we plot the last turbulent points
in the ($Re$, $R_\Omega$) space, for different resolutions and/or
codes on Fig.~\ref{graphcyc}. Turbulence is found on the cyclonic
side at least up to $R_\Omega=0.3$, i.e significantly away from the
marginal stability point.

Note that turbulence is maintained with certainty (with our
adopted criteria) at any given point, but, due to the sampling
made in the explored Reynolds number, turbulence may also be
maintained at a somewhat lower Reynolds number (i.e just below the
last turbulent point on Fig.~\ref{statcyc}). This can be true down
to the previously tested Reynolds number, for which turbulence is
not maintained at the considered rotation rate. In conclusion, the
real transition Reynolds $Rg$ curve in the ($R_\Omega, Re$) plane
should be found somewhat below (but not far from) the last
turbulent point curve determined here. This remark is more
important for anticyclonic flows, for which precise quantitative
results are needed.

Except for a systematic shift between the results obtained with
the Fourier code and the ZEUS-like one, the results seem to be
independent of the resolution. The numerically minded reader may
ask how one can reach such high Reynolds numbers with such
relatively small resolutions. This point is addressed in section
\ref{eff-re}.

\begin{figure}
  \centering
  \includegraphics[scale=0.35]{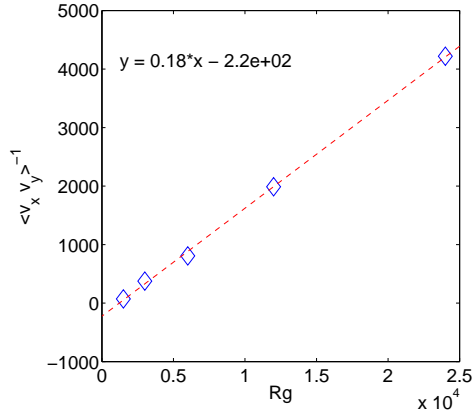}
  \caption{\small Mean transport as a function of the transition Reynolds number
   for cyclonic rotation (normalized by $S^2 d^2$). }\label{transcyc}
\end{figure}

An important issue is to quantify transport in subcritical
turbulent flows. The phenomenological arguments of \cite{L02}
suggest that $\langle v_x v_y\rangle\propto 1/Rg$ in subcritical
flows, and that the turbulent transport in a given flow with
specified $(Re, R_\Omega)$ numbers depends only on $R_\Omega$
through $Rg$ (see section \ref{phen})\footnote{The same
result follows if one assumes that in the fully turbulent state,
the torque $\propto Re^2$, as predicted in Kolmogorov turbulence,
and observed in experiments (see, e.g., \citealt{DDDLRZ05}). The
argument of section \ref{phen} allows us to recover this result
from more generic physical principles.}. Consequently, we have
used all our simulations at a given $R_\Omega$ to obtain the least
noisy evaluation of $\langle v_x v_y\rangle$. Then, with the help
of Fig.~\ref{statcyc}, one finds a transition Reynolds number $Rg$
for any given $R_\Omega$, which allows us to plot the mean
turbulent transport $\langle v_x\,v_y \rangle $ as a function of
the transition Reynolds number in Fig.~\ref{transcyc}. This was
done only from the data of our Fourier code for self-consistency,
but using both the $32^3$ and $64^3$ resolution runs, as they
produced the same results, and as the use of a larger data set
improves the statistics. The resulting relation reads

\begin{equation}\label{transp-cyc}
  \langle v_x\,v_y \rangle \simeq \frac{5.5}{Rg-1250}(Sd)^2.
\end{equation}

\noindent The presence of an additive constant in the denominator
of this expression is a clear indication of the influence of the
linear instability close to the marginal stability limit; indeed,
transport in the supercritical region is significantly enhanced
with respect to the subcritical region (see, e.g., Fig.~16 in
\citealt{DDDLRZ05}, and explanations therein). For large critical
Reynolds number (i.e., far enough from the marginal stability
boundary, e.g., $Rg\gtrsim 15000$), $\langle v_x\, v_y \rangle
\simeq 5.5/Rg$ is a good approximation.

\subsection{Numerical results: anticyclonic
flows}\label{res-anticyc}

\begin{figure}
  \centering
  \includegraphics[scale=0.45]{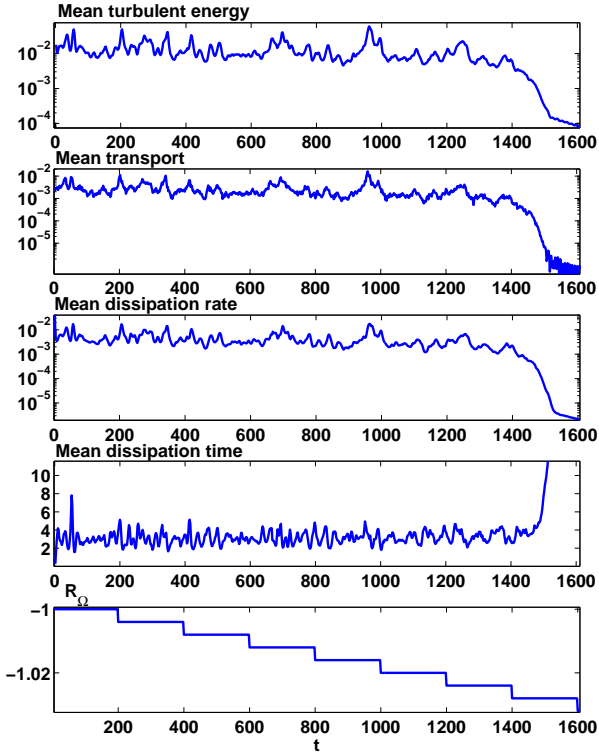}
  \caption{\small Time evolution of a $64^3$ Re=12000 anticyclonic flow as
   computed by our Fourier code. Panel description is identical to
   Fig.~\ref{statcyc}.}\label{statacyc}
\end{figure}

The strategy adopted in simulations of anticyclonic flows is
similar to the cyclonic side. Starting at $R_\Omega=-1.0$, the
rotation number is increased in steps of -0.004 and each step
lasts 200 shear times to allow for flow relaxation. A typical run
is shown on Fig.~\ref{statacyc}, computed with our 3D Fourier Code
at $Re=12000$. One should note that the flow fluctuations have
higher amplitudes on the anticyclonic side than on the cyclonic
side; this is why we have reduced the rotation number steps and
increased the relaxation time in anticyclonic runs. Consistently,
The last turbulent point is defined here as the last rotation rate
for which turbulence is sustained for $200 t_s$. On the example
Fig.~\ref{statacyc}, we find the last turbulent point for
$Re=12000$ at $R_\Omega=-1.024$.

\begin{figure}
  \centering
  \includegraphics[scale=0.35]{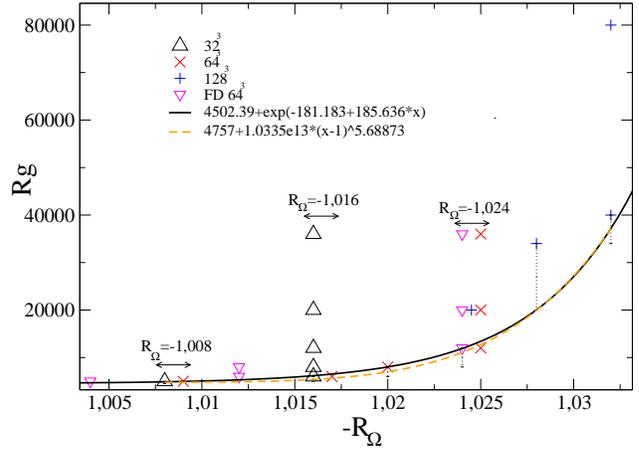}
  \caption{\small Transition Reynolds number $Rg$ as a function of the
  Rotation number $R_\Omega$, and related analytical fits, with different
  resolutions and codes for shearing-sheet boundary conditions (anticyclonic side).
  All plots were computed using our Fourier code except FD (finite difference)
  which uses our ZEUS-like code. Note that the x-axis is inverted with respect
  to Fig.~\ref{rotdata}. Symbols along the fitted lines correspond to
  resolved simulations; vertically aligned symbols indicate the limiting rotation
  number that can be reached at a given resolution and mostly correspond to unresolved
  simulations. For the sake of clarity, symbols which sit on top
  of each other have been slightly displaced along the $R_\Omega$
  axis; this is indicated by the arrows and the related values of
  $R_\Omega$.}\label{graphacyc}
\end{figure}

\begin{figure}
  \centering
  \includegraphics[scale=0.35]{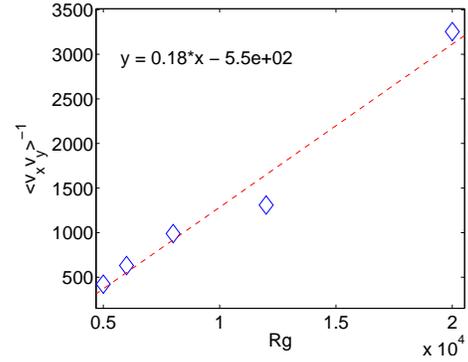}
  \caption{\small Mean transport as a function of critical reynolds number on
  the anticyclonic side (normalized by $S^2 d^2$).}\label{transacyc}
\end{figure}

As for cyclonic rotation, the last turbulent points for
anticyclonic rotation are plotted on Fig.~\ref{graphacyc} and the
mean transport on Fig.~\ref{transacyc}. Error bars are added on
Fig.~\ref{transacyc} to help assessing the significance of the
various fits performed, as they will be used later on. On the
lower bound of these bars, turbulence is not maintained with
certainty whereas the contrary is true for \emph{at least} 200
shear times at the upper bound. Therefore, the actual transition
Reynolds number is bracketed by the error bar.

Recalling that $R_\Omega=-2/q$ with $\Omega(r)\propto r^{-q}$ and
that $R_\Omega=-1$ corresponds to a constant specific angular
momentum distribution in cylindrical flows, the largest rotation
number reached here ($R_\Omega=-1.032$) corresponds to $q=1.94$;
this is quite consistent with the results shown on Fig.~1 of
\cite{HBW99}, except for the crucial fact that the resolution and
Reynolds number dependence are now quantified. The reason why such
high Reynolds numbers are accessible with our relatively low
resolutions is discussed in section \ref{eff-re}. For the time
being, let us comment a bit further on the information encoded in
Fig.~\ref{graphacyc}, which shows that Reynolds number and
resolution are different, albeit related control parameters. We
will focus on the Fourier code data for definiteness. Consider the
$32^3$ data, for example. For $|R_\Omega| < 1.016$, the transition
Reynolds number agrees with the one found at higher resolution.
However, increasing the Reynolds number above $\sim 6000$ produces
a loss of turbulence at the same rotation number independently of
the Reynolds number, whereas this is not true at higher
resolutions. This implies that the physics is not faithfully
represented at this resolution for $Re > 6000$ and $R_\Omega >
1.016$.

This is the most important point to note here: two different
regimes of transition from turbulent to laminar are displayed in
this figure. The first (corresponding to the various fitting
curves) is the correct, resolution independent and Reynolds
dependent transition. The second (apparent as the various
vertically aligned points at a given resolution) is an incorrect,
Reynolds independent and resolution limited transition. Note that
the points belonging to both this vertical line and the
laminar-turbulent line are still resolved, though, as shown in
section \ref{diss}. The meaning of the behavior displayed in
Fig.~\ref{graphacyc} is further discussed in section \ref{phen},
and its implications in sections \ref{kep} and \ref{eff-re}.

Comparing Figs.~\ref{graphcyc} and \ref{graphacyc}, we remark that
the dependence of the transition Reynolds number $Rg$ on the
``distance" to marginal stability in rotation number
$|R_\Omega-R_\Omega^{\pm}|$ is considerably stiffer on the
anticyclonic side than on the cyclonic one. This has important
implications that will be discussed in the next section.
Conversely, the turbulent momentum transport is very similar to
the one found for the cyclonic side\footnote{Fig.~\ref{transacyc}
is noisier than its cyclonic counterpart. This is a consequence of
the larger turbulent fluctuations observed in anticyclonic flows.
Longer integrations time-scale would have been required to improve
the statistics.}, as shown on Fig.~\ref{transacyc}
\begin{equation}\label{transp-acyc}
  \langle v_x\,v_y \rangle \simeq \frac{5.5}{Rg-3000}(Sd)^2.
\end{equation}

\noindent The constant in the denominator differs from the one
found on the cyclonic side. This reflects the difference of
transition Reynolds number at the two marginal stability limits.
For large enough Reynolds number, one find $\langle v_x\, v_y
\rangle \simeq 5.5/Rg$, which corresponds to the asymptotic
relation found on the cyclonic side (see section \ref{phen} for a
discussion of the possible origin of this behavior). This
indicates that this relation is very robust for subcritical flows,
far enough from the supercritical transition limit.

\section{Discussion}\label{discussion}

Our results are at variance with both the point of view
advocated by \cite{BHS96} and \cite{HBW99} (absence of subcritical
turbulence), and \cite{RZ99} and \cite{HDH05} (efficient transport
due to subcritical turbulence). This is further investigated in
this section. We shall first present some phenomenological
background material which helps to understand the physical origin
and meaning of the results presented in the previous section.
Then, we shall respectively discuss the implications of our
results for Keplerian flows (section \ref{kep}), the stabilizing
role of the Coriolis force in subcritical flows (section
\ref{cor}), and the relation between Reynolds number and
resolution (section \ref{eff-re}); these last two items have been
highly controversial in the past decade. Section \ref{eff-re} also
discusses the relation of these results with the scale-invariance
argument of \cite{B04}. Finally the influence of the nature of the
adopted boundary conditions and aspect ratio on our results is the
object of section \ref{BC}, as well as and their relation to fluid
dynamics experiments. Note also that the discussion of the
boundary conditions helps quantifying possible biases introduced
by the sheering sheet boundary conditions with respect to actual
disk physics. The reader interested only in the astrophysical
implications of our results may focus on section \ref{kep}.

\subsection{Some aspects of subcritical turbulence
phenomenology:}\label{phen}

The phenomenology of subcritical turbulence has been discussed in
\cite{L02} and \cite{LD05}. Some directly relevant aspects for our
present purpose are presented here (and clarified where needed).

Turbulent transport is often quantified in terms of a turbulent
viscosity. As this description has been criticized in the past, a
brief discussion of its use here might be useful. First, note that,
in scale-free systems such as the ones studied here (the only
scale present being the simulation box size), one can always
assume that

\begin{equation}\label{turb-vis}
\langle v_x v_y\rangle=\nu_t S,
\end{equation}

\noindent as this only amounts to defining a turbulent viscosity
$\nu_t$ such that this relation is satisfied. In any case, as the
source of turbulence is the shear, the Reynolds stress $\langle
v_x v_y\rangle$ \textit{must} be some function of the shear $S$,
which cancels when the shear cancels.

Now, $\nu_t$ has the dimension of a length times a velocity, so
that one must therefore have, in our simulations,

\begin{equation}\label{alpha}
\nu_t=\alpha S d^2,
\end{equation}

\noindent as $Sd$ and $d$ are the only dimensional quantities with
the right dimensionality introduced in the problem.

$\alpha$ is a Shakura-Sunyaev-like parameter. It is a
dimensionless quantity, and can therefore only depend on the
dimensionless quantities\footnote{Actually, in principle, $\alpha$
depends also on the aspect ratio of the simulation, and on the
nature of the boundary conditions. As these are not varied in the
results discussed on the basis of the phenomenology described
here, this dependency is ignored for simplicity.} characterizing
the problem at hand, namely the Reynolds number $Re$ and the
rotation number $R_\Omega$ (i.e., the shear dependence of $\alpha$
can only appear through the ratios of the shear time scale to the
viscous and the rotation time scales):

\begin{equation}\label{alpha-dep}
\alpha\equiv\alpha(Re, R_\Omega).
\end{equation}

 The results of section \ref{numerics} suggest that, quite remarkably,
$\alpha$ depends only on $R_\Omega$ through the transition
Reynolds $Rg$, and not (or little) on $Re$, in subcritical flows.
The origin of this behavior can be understood in the following way
(\citealt{L02}).

A sheared flow is out of global thermodynamical equilibrium, and
tries to restore this equilibrium by transporting momentum across
the shear. A subcritical flow has only two means at its disposal
to achieve this purpose: laminar and turbulent transport. It will
tend to choose the most efficient one under any given set of
conditions\footnote{Note that this does not imply that the
momentum transport is absolutely maximized.}, i.e. at given $Re$
and $R_\Omega$. The subcritical turbulent transport will exceed
the laminar one when $\nu_t \gtrsim \nu$. Right at the
laminar-turbulent threshold, $Re\sim Rg$ and $\nu_t\sim \nu$. This
implies that

\begin{equation}\label{alpha-Rom}
  \alpha\sim\frac{\nu}{Sd^2}\sim\frac{1}{Rg}.
\end{equation}

\begin{figure}[htb]
  \centering
  \includegraphics[scale=0.5]{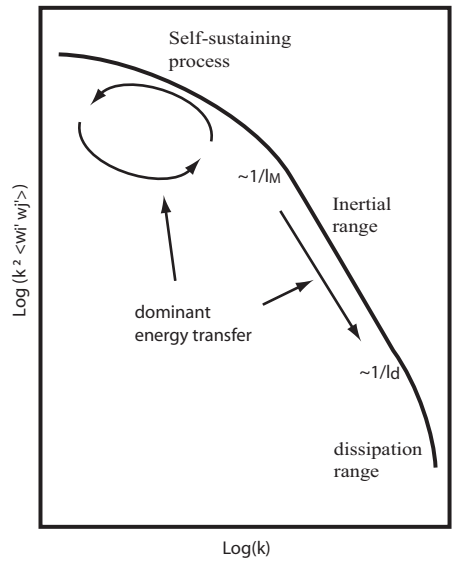}
  \caption{\small Proposed sketch of the idealized energy spectrum in a
  turbulent shear flow. Arrows indicate the energy flow through
  mode coupling. The length-scales $l_M$ and $l_d$ correspond to
  the top of the inertial range (assumed identical to the bottom
  of the self-sustaining range for simplicity) and the top of the
  dissipation range. Scales are assumed to be normalized to the box
  simulation size $d$, and anisotropy is ignored in this sketch
  (see text for details).}\label{spectrum}
\end{figure}

Now, what does happen at Reynolds numbers $Re$ \textit{larger}
than the transition Reynolds number $Rg$ ? To answer this
question, it is useful to have in mind some idealized, qualitative
picture of the situation in wave-number space. Such a picture is
proposed in Fig.~\ref{spectrum}, and constitutes a reasonable
working hypothesis. It is reasonably well-supported by our current
knowledge of the plane Couette flow turbulent self-sustaining
process and of inertial spectra, as well as by the spectral
analysis of some of our simulations presented and discussed in
section \ref{diss}.

In this picture, the large scales are occupied by the
self-sustaining mechanism. All scales in this domain are expected
to be coherent in phase, and interactions between large and small
scales occur both ways\footnote{This is the case in particular for
the non-rotating plane Couette self-sustaining mechanism
\citep{W97}.}. The intermediate range is the inertial range of
turbulence; scales have no phase coherence, energy cascades to
smaller scales at a constant rate, provided by the self-sustaining
mechanism (as part of the mode coupling taking place in the
self-sustaining mechanism range of scales occurs with the inertial
range). The smallest range represents the viscous dissipation
scales. The existence of the self-sustaining process scales, their
properties, and their influence on the inertial range (energy
input and anisotropy) is the distinctive feature of shear
turbulence with respect to the more commonly known and studied
forced isotropic turbulence.

In such a picture, increasing the Reynolds number almost
exclusively results in an increase of the inertial range, which is
essentially vanishing at the transition Reynolds number. This
should have little effect on the turbulent transport (whereas, on
the contrary, the laminar transport becomes smaller and smaller
when increasing the Reynolds number).

Indeed, we have first checked that this is case in non-rotating
Couette plane flows, where the self-sustaining mechanism is
identified \citep{HKW95}: the transport is almost completely
determined dominated by the mechanism rolls and streaks.
Furthermore, in our simulations, we have computed the contribution
of each length scale to the total transport $\langle v_x v_y
\rangle$. First one should note that in Fourier space (in 1D for
simplicity):

\begin{equation}
\langle v_x v_y \rangle=\sum_{n=0}^{N-1} \tilde{v_x}(k_n) \tilde{v_y^*}(k_n)
\end{equation}

\begin{figure}[htb]
  \centering
  \includegraphics[scale=0.4]{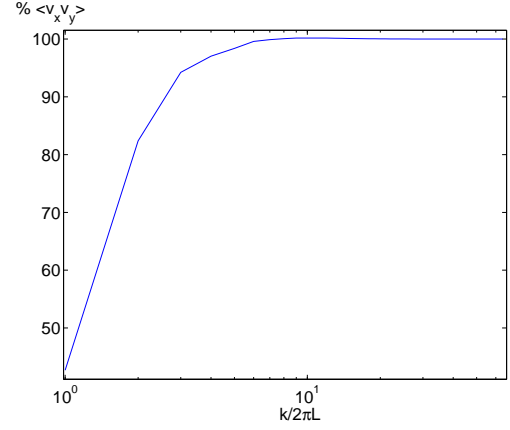}
  \caption{\small Example of the cumulative sum contribution of each scale length to the mean transport,
  starting from $k_x=0$ in the x direction: 99\% of the transport comes from $k_x<10$ scales.
  From a $128^3$ simulation, Re=20000,
  $R_\Omega=-1.020$. }\label{trans_spec}
\end{figure}

\noindent Therefore, the contribution to mean transport of the
wavelength $k_n$ is found to be $2\Re
\Big(\tilde{v_x}(k_n)\tilde{v_y^*}(k_n)\Big)$, since $k_n$ and
$k_{N-n}$ represents the same physical wavenumber. This simple
result can be used in 3D by averaging the transport over 2
directions (in physical space) and by computing the Fourier
transform in the remaining direction ; this procedure is
sufficient for our purpose here. The resulting cumulative Fourier
sum, starting from $n=0$ is illustrated on an example on
Fig.~\ref{trans_spec} to quantify which scales dominate the
transport. In this example, the resulting spectral analysis is
plotted in the $x$ direction for a $128^3$ anticyclonic flow with
$R_\Omega=-1.024$ and $Re=20000$, showing that more than $99\%
$
of the transport comes from scales larger than 1/10 of the box
size ; this range corresponds to the length-scales of the
self-sustaining process (see section \ref{diss}). Similar results
are found for spectral analyzes in the $y$ and $z$ directions,
consistently with the picture discussed here. This is expected
anyway if the inertial spectrum is kolmogorovian, as confirmed
from the spectral analysis of section\footnote{The true nature of
the inertial spectrum might be affected by the anisotropy
generated by the the shear and the Coriolis force, but these
anisotropies must become negligible at small scale, due to the
shorter and shorter eddy turnover time.} \ref{diss}.

There are two loose ends in this discussion. First, hysteresis is
usually experimentally observed in subcritical transitions to
turbulence: the measured transition Reynolds number is higher when
moving ``up" from the laminar to turbulent states than when moving
``down" from the turbulent to laminar ones. This suggests that the
laminar-turbulent boundary is separated by some sort of barrier in
the appropriate phase-space (defined, e.g., by the amplitudes and
phases of the Fourier modes). This (along with the fact that the
arguments developed here apply only in order of magnitude) may
well explain the existence of the constant of order $5$ that one
finds in Eqs.~(\ref{transp-cyc}) and (\ref{transp-acyc}) with
respect to Eq.~(\ref{alpha-Rom}). Secondly, the arguments
presented here ignore the existence of marginal stability
thresholds. This, as pointed out in sections \ref{res-cyc} and
\ref{res-anticyc}, may explain the presence of the constant at the
denominator of these relations, as the equivalent global
subcritical transition Reynolds number that one can define in the
supercritical regime is orders of magnitude smaller than in the
subcritical regime.

To conclude, let us point out the relation of this picture with
the numerical results presented in Fig.~\ref{graphacyc}. The fact
that higher resolutions are required to faithfully represent the
physics at higher rotation numbers indicates that the ratio
$d/l_M$ increases with rotation number. Indeed, if the resolution
is too low, so that the relative scale $l_M/d$ is not resolved,
the energy transfer loop represented on Fig.~\ref{spectrum} cannot
take place, and turbulence is not self-sustained. Furthermore, at
the transition Reynolds number, the inertial spectrum is nearly
inexistent, as pointed out above, and $l_M\sim l_d$. Consequently,
the most critical scale ratio in this problem is expected not to
be the Kolmogorov one, but the self-sustaining mechanism one
($d/l_M$).

\subsection{Implications for Keplerian flows\label{kep}}

Actual disks are vertically stratified, whereas stratification is
ignored in our experiments. Stratification provides us with a
local macroscopic scale (the disk scale height $H$). With
appropriate provisos related to the possible stabilizing or
destabilizing role of stratification\footnote{If stratification is
destabilizing, the momentum transport induced by the resulting
convective motions is in the wrong direction, as recalled in the
introduction, and must be counterbalanced by another process;
ignoring stratification in this case therefore makes life easier
for this other process (here, subcritical turbulence). If
stratification is stabilizing, this also most likely results in an
increased difficulty in finding the transition to turbulence, and
a related increase in the transition Reynolds number. These
arguments suggest that ignoring the dynamical stratification
altogether maximizes the overall outwards transport in our
problem.}, one can tentatively identify this scale height with our
simulation box size: $H=d$. This assumption is made throughout
this section. In the same way, the Shakura-Sunyaev $\alpha_{SS}$
parameter is defined such that $\nu_t= \alpha_{SS} c_s H\simeq
\alpha_{SS}\Omega H^2$. Eq.~(\ref{alpha}) then implies that
$\alpha_{SS}=2\alpha/|R_\Omega|\simeq\alpha$ (the last equality
holds within a factor of order unity for the rotation number range
of interest in this work).

Using the numerical results shown on Figs.~\ref{graphacyc} and
\ref{transacyc}, one can deduce a few properties of Keplerian flow
subcritical shear turbulence, based on various conservative
extrapolations of our numerical data. First, the transition
Reynolds number $Rg$ dependence on the rotation number $R_\Omega$
is well-fitted by a power or an exponential law. Using these laws,
one can get a first set of estimates of the transition Reynolds
number for keplerian-like flows ($R_\Omega=-4/3$): $Rg=1.1\times
10^{10}$ and $Rg=1.3 \times 10^{26}$, respectively. The last
estimate leads to the absence of subcritical turbulence in
accretion disks whereas the first one allows for its
existence\footnote{We assume that accretion disk Reynolds numbers
lie between $10^{10}$ and $10^{15}$ for definiteness. The Reynolds
number definition used in this evaluation is $Re=SH^2/\nu$ where
$H$ is the local disk scale height, consistently with the $H=d$
identification made earlier.}. Secondly, let us note that, for
both cyclonicity, the Coriolis force induces a steeper and steeper
increase of the transition Reynolds number when moving away from
the marginal stability boundary. This suggests that one can find a
lower bound for $Rg$ by linearly extrapolating the power law fit
beyond the last known point ($R_\Omega=-1.032$). One find this way
$Rg^{min}=1.8\times 10^6$. As a final hypothesis, one may envision
that the $Rg(R_\Omega)$ relation would be more or less symmetric
with respect to $R_\Omega=0$ if there were no supercritical
domain. This would explain why the actual relation of
Fig.~\ref{graphacyc} is so steep: in this picture, the system
tries to reach back as fast as possible the high values of
transition Reynolds number expected from this hypothetical
symmetry, after which the Reynolds dependence with rotation number
would be much less steep. Under this assumption the expected
transition Reynolds number for keplerian flows would be
$Rg=2.\times 10^7$ (a power-law fit of the cyclonic data has been
used in this extrapolation).

\begin{table}
  \centering
  \caption{Extrapolated transition Reynolds numbers, values of $\alpha$, and required simulations
  resolution, for keplerian flows, under various assumptions (see text for details)}\label{kep-tab}
\begin{center}
\begin{tabular}{lllll} \hline
  % after \\ : \hline or \cline{col1-col2} \cline{col3-col4} ...
   & exponential & power-law & cyclonic & linear \\ \hline
  Rg & $1.3\times 10^{26}$ & $1.1\times 10^8$ & $2\times 10^7$ & $1.8\times 10^6$ \\
  $\alpha$ & n/a & $5\times 10^{-10}$ & $2.6\times 10^{-7}$ & $3.1\times 10^{-6}$ \\
  $(d/\delta)^3$ & n/a  & $7000^3$ & $3000^3$ & $900^3$ \\ \hline
\end{tabular}
\end{center}
\end{table}

% linear law is: Re=37000+(rom-1.032)*6.17e6}

This information is summarized in table \ref{kep-tab}, along with
the corresponding values of $\alpha$, obtained from the asymptotic
relation $\alpha=\langle v_x v_y \rangle =5.5/Rg$ found for
cyclonic and anticyclonic flows in the previous section. The last
line shows the resolution required to successfully simulate
keplerian flow turbulence, for the various Reynolds numbers (see
subsection \ref{reso}). One sees that even the most optimistic
$\alpha$ bound ($\alpha_{max}=3.1\times 10^{-6}$), obtained with
the linear extrapolation, is substantially smaller than the values
required in astrophysical accretion disks (as summarized, e.g., in
\citealt{PL95}). Note finally that, even without any
extrapolation, our results exclude subcritical turbulent transport
at the $\alpha\simeq 3.10^{-4}$ level.

\subsection{Role of the Coriolis force in uniform shear
flows}\label{cor}

Two different but related issues have been raised in the
literature concerning the role of the Coriolis force in
subcritical systems.

First, for linearly stable flows, \cite{BHS96} point out that the
Coriolis force plays a conflicting role in Eqs.~(\ref{wx2av}) and
(\ref{wy2av}). More precisely, they make the following point: as
$S\langle v_x v_y\rangle > 0$ for turbulence to exist (see section
\ref{energy}), the terms in which the shear $S$ has been factored
out in these equations have opposite signs for linearly stable
flows, while they have the same sign for linearly unstable flows
(note that this is true independently of the flow cyclonicity).
They conclude from this that a stabilizing rotation prevents
turbulence to show up in subcritical shear flows, except possibly
in the vicinity of marginal stability. Somewhat relatedly, the
recent Reynolds stress-closure model of \cite{O03} and \cite{GO05}
predicts relaminarization for large enough deviations from the
marginal stability limit. In particular, for the authors' standard
choice of parameters, it predicts relaminarization for
$R_\Omega\sim 0.2$ for cyclonic rotation. However, as can be seen
on Fig.~\ref{graphcyc}, both the \cite{BHS96} argument and the
\cite{GO05} result conflict with our simulations: subcritical
turbulence is maintained away from marginal stability on the
cyclonic side, at least up to $R_\Omega \simeq 0.3$. Note that we
could have pushed the search for transition to turbulence beyond
what is shown on this graph, especially by using higher
resolutions, but did not do it due to computer resources
limitations. As discussed in the next subsection, the absence of
turbulence in the keplerian flow simulations of \cite{BHS96} and
\cite{HBW99} is a problem of resolution.

The second issue relates to the asymmetry between cyclonic and
anticyclonic rotation. The stress-closure model just mentioned
depends on the rotation number only through the combination
$R_\Omega(R_\Omega+1)$ which implies a symmetry with respect to
$R_\Omega=-1/2$. This symmetry is clearly violated by our
numerical results (compare Figs.~\ref{graphcyc} and
\ref{graphacyc}), a point which requires some comments.

First, note that the linearized Navier-Stokes equation
(\ref{split-RPC}) exhibits this symmetry for perturbations with
vanishing pressure variation ($\delta\pi=0$). In this case, the
linearized equation can be written:

\begin{equation}\label{eqnlin}
  \frac{\partial{\mathbf w}}{\partial t}
 =
  S\cdot y \frac{\partial{\mathbf w}}{\partial x}
   +  S\cdot \Big((R_\Omega+1) w_y {\mathbf e}_x
  - R_\Omega w_x {\mathbf e}_y\Big)
    +\nu\Delta{\mathbf w}
\end{equation}

\noindent The cyclonic-anticyclonic symmetry appears when
exchanging the $x$ and $y$ directions. Indeed, upon the following
change of variables:

\begin{eqnarray}
R'_\Omega&=-R_\Omega-1, \nonumber \\
w'_x&=w_y,\qquad&{\mathbf e'}_x={\mathbf e}_y,\nonumber \\
w'_y&=w_x,\qquad&{\mathbf e'}_y={\mathbf e}_x,\nonumber \\
w'_z&=w_z,\qquad&{\mathbf e'}_z={\mathbf e}_z,\nonumber
\end{eqnarray}
\noindent so that
\begin{eqnarray}
 \mathbf w'&=&w'_x {\mathbf e'}_x+w'_y {\mathbf e'}_y+w'_z {\mathbf e'}_z\nonumber\\
          &=&\mathbf w,\nonumber
\end{eqnarray}

\noindent the form of Eq.~(\ref{eqnlin}) should be invariant,
which is indeed the case:

\begin{equation}\label{eqnlinsym}
  \frac{\partial{\mathbf w'}}{\partial t}
 =
  S\cdot y \frac{\partial{\mathbf w'}}{\partial x}
   + S\cdot \Big((R'_\Omega+1) w'_y {\mathbf e'}_x
  - R'_\Omega w'_x {\mathbf e}_y\Big)
    +\nu\Delta{\mathbf w'},
\end{equation}

\noindent This symmetry can also be extended to compressible
motions by adding $\delta\pi'(x,y,z)=\delta\pi(y,x,z)$ to the list
of change of variables.

Because the perturbations defining the linear stability limit also
exhibit this symmetry (Appendix \ref{app:disp}), it has often been
assumed in closure-stress models in the past. However, this is not
a symmetry of the full Navier-Stokes equation
\citep{SMM89,S91,SC97}, nor of the ${\bm\nabla}\cdot{\bm w}=0$
equation). This is also apparent in a direct inspection of the
structure of simulated turbulent flows. The $R_\Omega=0$,
wall-bounded turbulent flows contain large \textit{streamwise}
rolls living for about a hundred shear times \citep{HKW95}. We
have also found rolls more or less aligned in the streamwise
direction in our $R_\Omega=0$ shearing sheet simulations, although
we did not try to precisely quantify their survival time.
Furthermore, at the anticyclonic marginal stability limit
($R_\Omega=-1$), we did observe sheared \textit{shearwise} rolls
(i.e rolls in $y$ direction) in our simulations, as one might
expect from the symmetry of the linearized Navier-Stokes equation.
The anticyclonic roll survival time is observed to be rather short
compared to their cyclonic counterpart, as they are tilted by the
shear and loose their coherence in a few shear times at most. This
roll lifetime is the main difference we found between the cyclonic
and anticyclonic side. This is related to the fact that a
streamwise roll does not reduce the shear on the anticyclonic
subcritical domain (in opposition to the cyclonic one).

In any case, we have found turbulence away from the marginal
stability limit in cyclonic flows, and the symmetry with respect
to $R_\Omega=-1/2$ is violated both in our simulations, and in
supercritically rotating shear flow turbulence (see, e.g.,
\citealt{SC97} and references therein). This make the predictions
of the stress-closure model of \cite{O03} and \cite{GO05}
unreliable in both subcritical and supercritical flows.

\subsection{Resolution, effective Reynolds number and scale invariance\label{eff-re}}

The results of section \ref{res-cyc} and \ref{res-anticyc} involve
fairly high Reynolds numbers, and one might ask if our simulations
are resolved enough in these regimes. This question has a priori
two different aspects, as one can guess from Fig.~\ref{spectrum}:
resolving the self-sustaining process smallest relative scale
$d/l_M$, and resolving the relative dissipation scale $d/l_d$.

For the problem considered in this paper, resolving the first
scale is a \textit{sine qua non} condition: if it is not
satisfied, turbulence does not show up, independently of the
simulation Reynolds number, because the required scale coupling
shown on Fig.~\ref{spectrum} for the self-sustaining process to
exist cannot take place. This shows up in Fig.~\ref{graphacyc} as
the vertical transition limit from turbulent to laminar that we
obtained for any given resolution, for large enough Reynolds
numbers.

Resolving the dissipation scale is important to ascertain that
direct numerical simulations such as the ones performed here are
not biased by (the presence or absence of) numerical dissipation,
and this issue is often raised in the fluid mechanics literature.
For the time being, we note that, at the transition Reynolds
number, the inertial domain should be non-existent or extremely
reduced, so that $l_M\simeq l_d$ and both resolution requirements
should be directly related (this point, used in section \ref{reso}
is justified in section \ref{diss}). We can therefore consider that the
``effective Reynolds number" $Re_{\textrm{eff}}$ of our simulations is the
largest transition Reynolds number $Rg$ correctly determined at a
given resolution\footnote{With all the provisos discussed in section
\ref{init} about the role of the choice of the initial conditions and
turbulence minimal survival life-time.}, as discussed in section
\ref{res-anticyc}.

Note that this effective Reynolds number is problem-dependent: the
self-sustaining process qualitative and quantitative
characteristics both depend on the considered problem;
furthermore, in simulations of isotropic turbulence, the
self-sustaining process is absent, and replaced by a forced
amplitude of the largest Fourier modes, so that the effective
Reynolds number in this case is the one related to the dissipation
scale.

Let us now examine the two requirements mentioned above in more
detail.

\subsubsection{Resolving the self-sustaining process}\label{reso}

First, we would like to qualitatively comment on the difference of
resolution requirements between cyclonic and anticyclonic flows.

As discussed in section \ref{BC}, the nature of the shearwise
boundary condition has apparently only a small influence on the
results; this is exemplified by the similar transition Reynolds
numbers found in our simulations and in experiments on rotating
shear flows (see Fig.~\ref{TA-num}). This suggests that at least
some of the characteristics of the self-sustaining process of
non-rotating plane Couette flows are relevant here. At the
cyclonic marginal stability limit, this self-sustaining process
has a time-scale $t_{\textrm{SSP}}\sim 100 S^{-1}$
\citep{HKW95,W97}. The requirement that, at the transition
Reynolds number, the viscous time scale at scale $l_M$ exceeds
$t_{\textrm{SSP}}$ reads $l_M^2/\nu\gtrsim 100 S^{-1}$, i.e.,
$l_M/d\lesssim (100/Rg)^{1/2}\sim 1/4$ for $Rg\sim 1500$
\citep{LD05}. This probably explains why the resolution
requirement is so low on the cyclonic side. Conversely, we have
mentioned at the end of the previous subsection that rolls (which
are an apparently ubiquitous ingredient in subcritical turbulence)
do not survive more than a few shear times in anticyclonic flows.
Therefore, the anticyclonic self-sustaining process time-scale
cannot exceed a few shear times as well, whatever its nature. The
same reasoning as the one exposed above leads to $l_M/d\lesssim$ a
few $(1/Rg)^{1/2}\sim$ a few $\times 1/70$, an already much more
demanding constraint. It is obviously related to the larger
transition Reynolds number found at the anticyclonic marginal
stability, compared to the cyclonic one.

As mentioned several times already, the self-sustaining process is
identified and understood only at the cyclonic marginal stability
limit in wall-bounded Couette flows. Consequently, it is difficult
to explain why the resolution demand grows so much faster with
rotation number ``distance" to marginal stability for anticyclonic
flows than for cyclonic ones. However, we speculate that this is
connected to the fact the rotation time scale is only a fraction
of $S^{-1}$ for cyclonic flows, whereas it always exceeds $S^{-1}$
for anticyclonic ones.

Next, let us try to quantify the resolution that would be needed
to successfully simulate keplerian flows. The phenomenology of
subcritical turbulence developed by \cite{L02} predicts that
$d/l_M\sim Rg^{1/2}$ and $\langle v_x v_y\rangle \propto 1/Rg$.
This phenomenology implicitly assumes that the relevant time-scale
of the self-sustaining process is $\sim S^{-1}$, so that it would
need to be modified to be applied to cyclonic flows, but it should
be adequate for anticyclonic ones, with appropriate modifications.
In particular, we have already pointed out in section \ref{phen}
that the last relation needs to be amended into $\langle v_x
v_y\rangle \propto 1/(Rg-R_c)$ with $R_c\simeq 3000$ on the
anticyclonic side. This suggests that

\begin{equation}\label{resol}
  \frac{\delta}{d}\simeq \frac{\gamma}{(Rg-R_c)^{1/2}}
\end{equation}

\noindent is the appropriately generalized scale relation
($\delta$ being the smallest scale accessible to the simulation,
i.e., the resolution). Table \ref{RRG} gives the values of
$\gamma$ and $Re_{\textrm{eff}}$ for the three different resolutions of our
anticyclonic simulations.

\begin{table}
\centering
\begin{tabular}{lll}
\hline
$(d/\delta)^3$ & $Re_{\textrm{eff}}$ &  $\gamma$ \\
\hline
$32^3$     & 6000       &    1.71    \\
$64^3$     & 12000      &    1.48    \\
$128^3$    & 38000      &    1.46    \\
\hline
\end{tabular}
\caption{\small Resolution, effective Reynolds number and $\gamma$
factor for the Fourier code on the anticyclonic side}\label{RRG}
\end{table}

Although the statistics is a little poor to draw firm conclusions,
it appears that $\gamma$ is nearly constant compared to the
variations in both resolution and transition Reynolds number, and
our simulations are therefore consistent with Eq.~(\ref{resol}).
The resolution needed to simulate keplerian flows has been
computed based on the estimate Eq.~(\ref{resol}), with
$\gamma=1.5$ (the $R_c$ correction has little influence on these
estimates). The results are shown in table \ref{kep-tab}. For
comparison purposes, note that the largest turbulence simulation
ever performed was $4000^3$, but was not run for hundreds or
thousands of dynamical times. Although the results gathered here
are probably only indicative, as they are based on guess work,
they strongly suggest that simulating subcritical turbulence in
keplerian flows is beyond present day computer capabilities, and
support the idea that the subcritical keplerian flows simulations
performed to date were limited by numerical resolution, as
suggested by \citet{L02}.

\subsubsection{Resolving the dissipation scale}\label{diss}

In statistically steady turbulence, the dissipation scale can be
defined from the balance between input and dissipation described
by Eq.~(\ref{w2avb}). The energy input is provided by $S\langle
v_x v_y\rangle$. The Fourier analysis of this quantity is shown on
Fig.~\ref{trans_spec}, and is dominated by the large scales.
Conversely, the Fourier content of $\epsilon$,
Eq.~(\ref{epsilon}), is dominated by the small scales (large $k$),
comparable to the dissipation scale, as illustrated below.

Resolving the dissipation scale is important with Fourier codes in
order to prevent energy accumulation at the smallest scales, which
may bias the results, or lead to code crash\footnote{One may also
include an hyper-viscosity term to prevent code crash, but this
turned out not to be necessary.}.

The general definition of the dissipation wavelength $k_d$ follows
from the evaluation of Eq.~(\ref{epsilon}) in Fourier space:

\begin{equation}
\label{ldef} \epsilon=2\nu\int_0^{k_d}k^2 E(k) dk
\end{equation}

\noindent where it is assumed that $E(k)$ is cut-off at $k_d$,
either abruptly, or through some modelling of the dissipation
range (see e.g. \citealt{L90}).

In simulations of homogeneous and isotropic turbulence, the energy
input is imposed from the outside: the amplitude of the largest
Fourier mode is held fixed, and Fig.~\ref{spectrum} reduces to the
inertial and dissipation range. In this context, the inertial
spectrum reduces to the Kolmogorov spectrum given by:

\begin{equation}
\label{ekdef} E_K(k)=C_K \epsilon^{2/3}k^{-5/3},
\end{equation}

\noindent where the Kolmogorov constant $C_K\simeq 1$. Cutting off
this spectrum at wavelength $k_d$ and injecting it in the
definition Eq.~(\ref{ldef}) leads to the well-known expression of
the Kolmogorov wave number, $k_K=(\epsilon/\nu^3)^{1/4}$. The
related Kolmogorov scale (inverse of the wave number) is a largely
used estimate of the dissipation scale.

In the fluid mechanics community one often requires that the
Kolmogorov scale be resolved, even if the considered turbulence is
not isotropic and homogeneous, as, e.g., in shear flow turbulence
(see, e.g., \citealt{P96}). However, in our simulations, the
observed spectrum is substantially different from the Kolmogorov
one, especially at the transition Reynolds number (see top panel
of Fig.~\ref{spectres}). Indeed, at the turbulent-laminar
transition, one does not expect nor observe the presence of an
inertial domain in the spectrum. One may therefore ask what
relation the Kolmogorov scale bears to the dissipation scale of
the problem.

Consider, e.g., the $32^3$ and $64^3$ energy spectra obtained at a
Reynolds number $Re=6000$ and a rotation number set to $-1.016$.
These spectra are shown on the top panel of Fig.~\ref{spectres}.
The concordance of the spectra at both $32^3$ and $64^3$
resolutions indicates that the dissipation scale in the $32^3$
simulation is resolved (this is consistent with the shape of the
spectrum at the smallest $32^3$ resolved scales, much steeper than
Kolmogorov). It appears that the largest distance to marginal
stability $|R_\Omega+1|$ reliably accessible at a given resolution
on the laminar-turbulent transition (as checked by higher
resolution simulations) corresponds to the various vertical line
of transition displayed on Fig.~\ref{graphacyc} for this
resolution. In other words, the $Re=6000$, $R_\Omega=-1.016$ point
at $32^2$, and the $Re=12000$, $R_\Omega=-1.024$ at $64^3$, are
resolved. This feature makes us confident that the transition
point determined at $128^3$ is the correct one, although we did
not cross-check it at $256^3$, due to the limitations in the
available computational resources.

We have thus determined the largest transition Reynolds number
where the dissipation scale is confidently resolved in these
anticyclonic runs at the various resolutions we have used ($32^3$,
$64^3$ and $128^3$). In other words, we know the effective
dissipation scale of these simulations, as it must be comparable
to the largest wave number available in the
simulation\footnote{This expression corrects a misprint in
\cite{P96} for the diagonal of a cube in Fourier space; although
this largest wave number is resolved only in discrete directions,
this definition is adopted here for ease of comparison with this
earlier work.}: $k_d\simeq 3^{1/2}\pi N/d$, where $N$ is the
resolution. Furthermore, we can compute the Kolmogorov wave number
$k_K$ for these runs, as $\nu=Re/Sd^2$, and as $\epsilon$ follows
from Eq.~(\ref{w2avb}) and the transport (e.g., with the help of
the transport/transition-Reynolds-number correlation displayed on
Fig.~\ref{transacyc}). The resulting ratio $R=k_d/k_K$ is given in
table \ref{ratio}.

\begin{table}
\centering
\begin{tabular}{lll}
\hline
$N$ & $R=k_d/k_K$ & $Rg$ \\
\hline
$32$     & 1.23       &    6000    \\
$64$     & 1.73       &    12000   \\
$128$    & 2.66       &    35000   \\
\hline
\end{tabular}
\caption{\small Resolution, dissipation to Kolmogorov wave number
ratio, and corresponding transition Reynolds number (see text for
details).}\label{ratio}
\end{table}

Although the values of the ratio $R$ quoted in table \ref{ratio}
are of order unity, a systematic trend seems to appear, indicating
that resolving the Kolmogorov wave number is possibly not the
relevant concept at the transition Reynolds number, as it is not
stringent enough; nevertheless, the required resolution derived
from the Kolmogorov wave number is apparently semi-quantitatively
correct, at least for the rotation numbers explored here. Of
course when going to Reynolds numbers well in excess of $Rg$ at a
given $R_\Omega$, the Kolmogorov wave number should always give
the right estimate of the dissipation scale, as the inertial range
becomes more and more prominent in the overall spectrum.

To conclude this aspect of the discussion, we note that both the
non-kolmogorovian shape of the spectrum at transition and the
relatively small values of $\epsilon$ at the various transition
Reynolds numbers used here, most probably combine in the end to
explain why we can reach rather large Reynolds numbers at rather
moderate resolutions.

In order to have a better grasp on which scales contribute most to
the dissipation, we have computed a quantity, $\tau_d(k)$, defined
by

\begin{equation}
\tau_d(k)=2\nu \int_{-k}^k \!\! dk_x\!\! \int_{-k}^k \!\! dk_y\!\!
\int_{-k}^k \!\! dk_z \,(k_x^2+k_y^2+k_z^2)E(k_x,k_y,k_z).
\end{equation}

\begin{figure}
   \centering
   \includegraphics[scale=0.4]{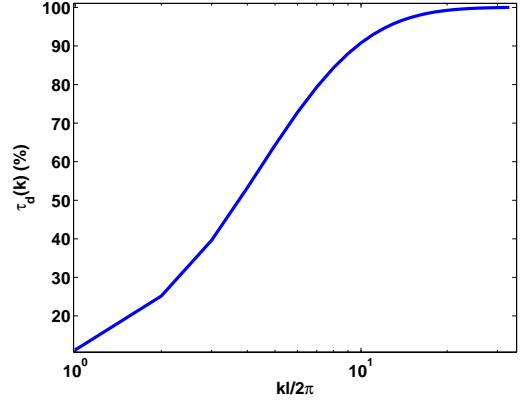}
   \caption{\small Cumulative mean dissipation spectrum for a $64^3$ simulation at Re=6000 for
   $R_\Omega=-1.016$.}{}\label{spectre_dissip}
\end{figure}

\begin{figure}
   \centering
   \includegraphics[scale=0.40]{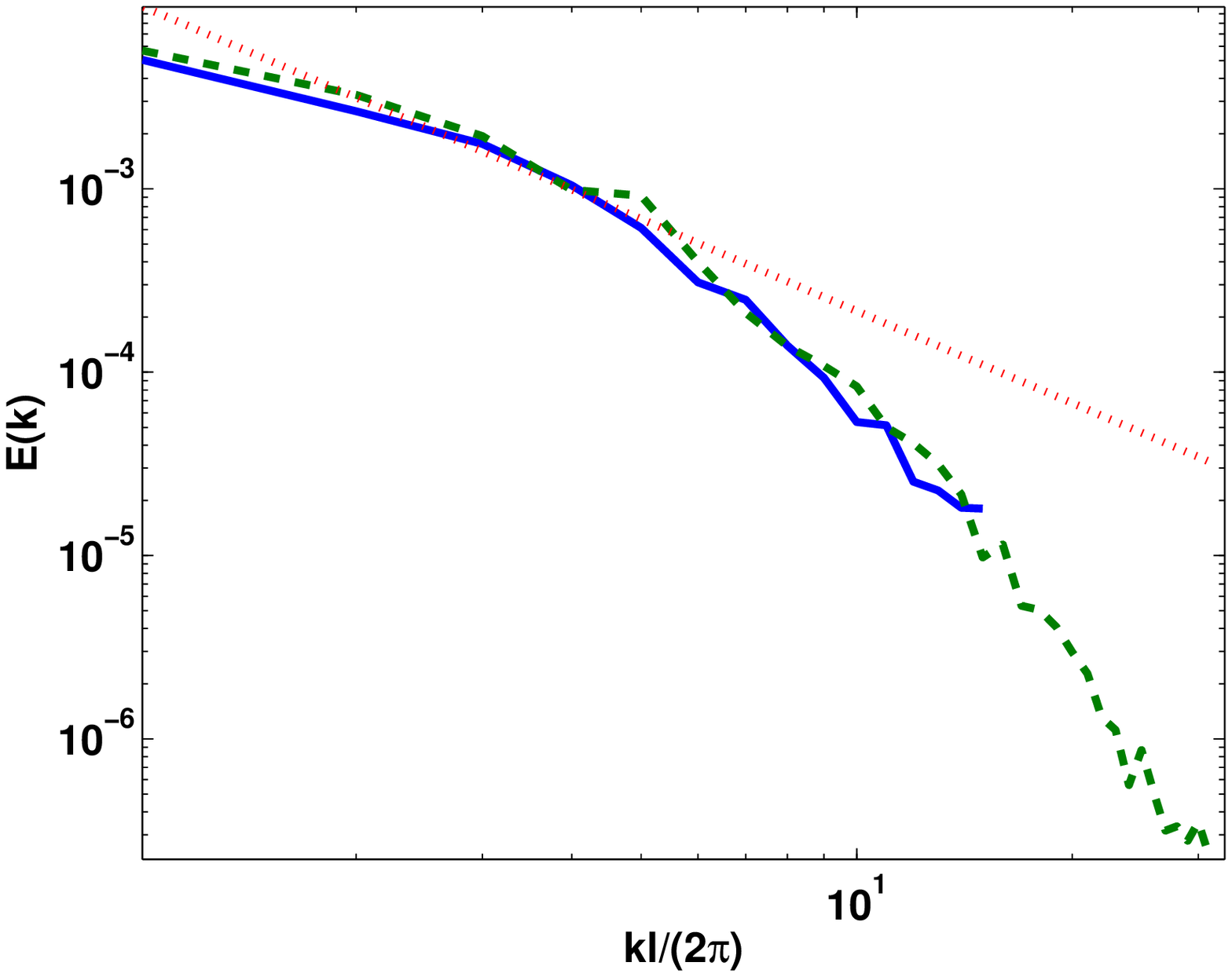}
   \includegraphics[scale=0.40]{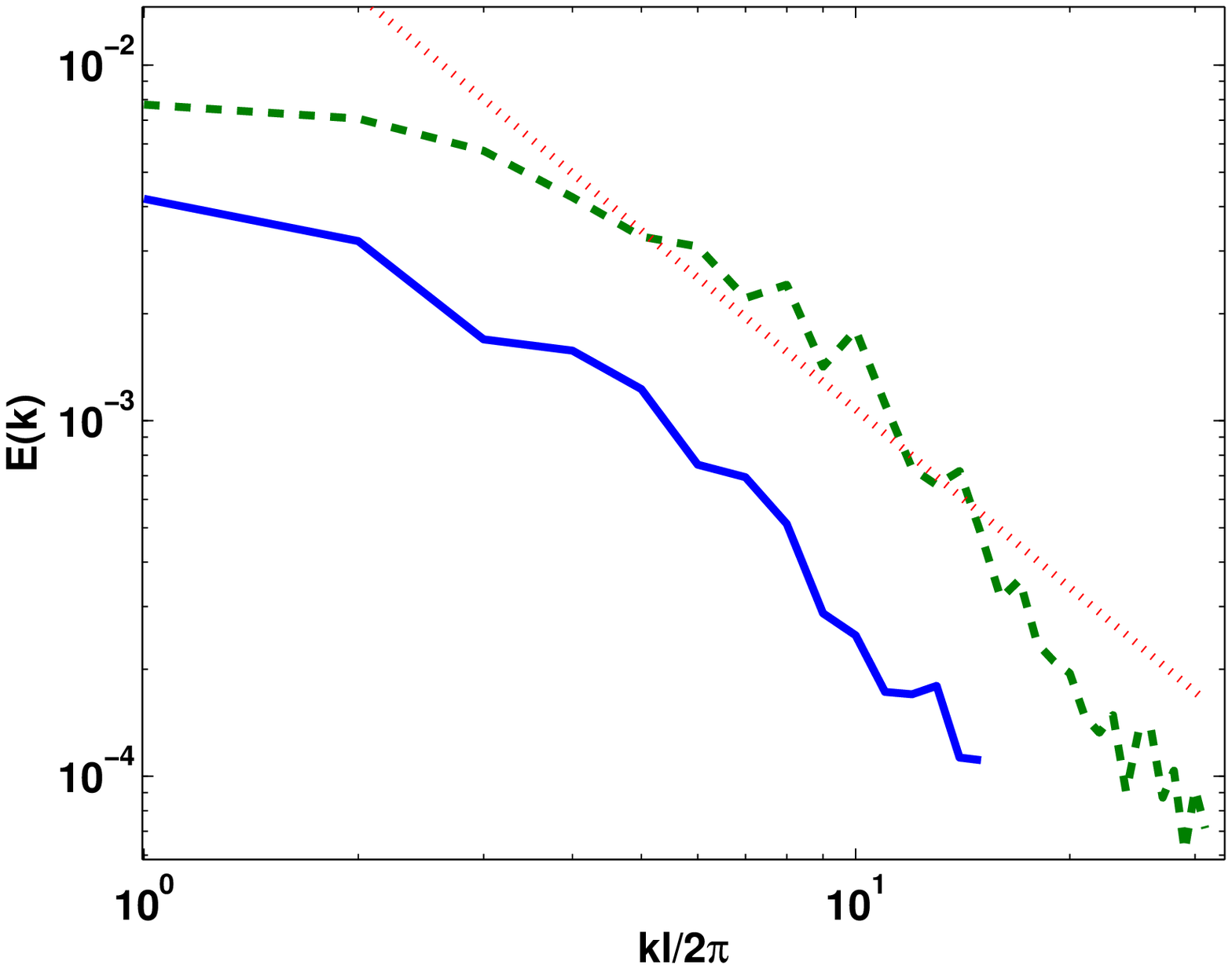}
   \includegraphics[scale=0.40]{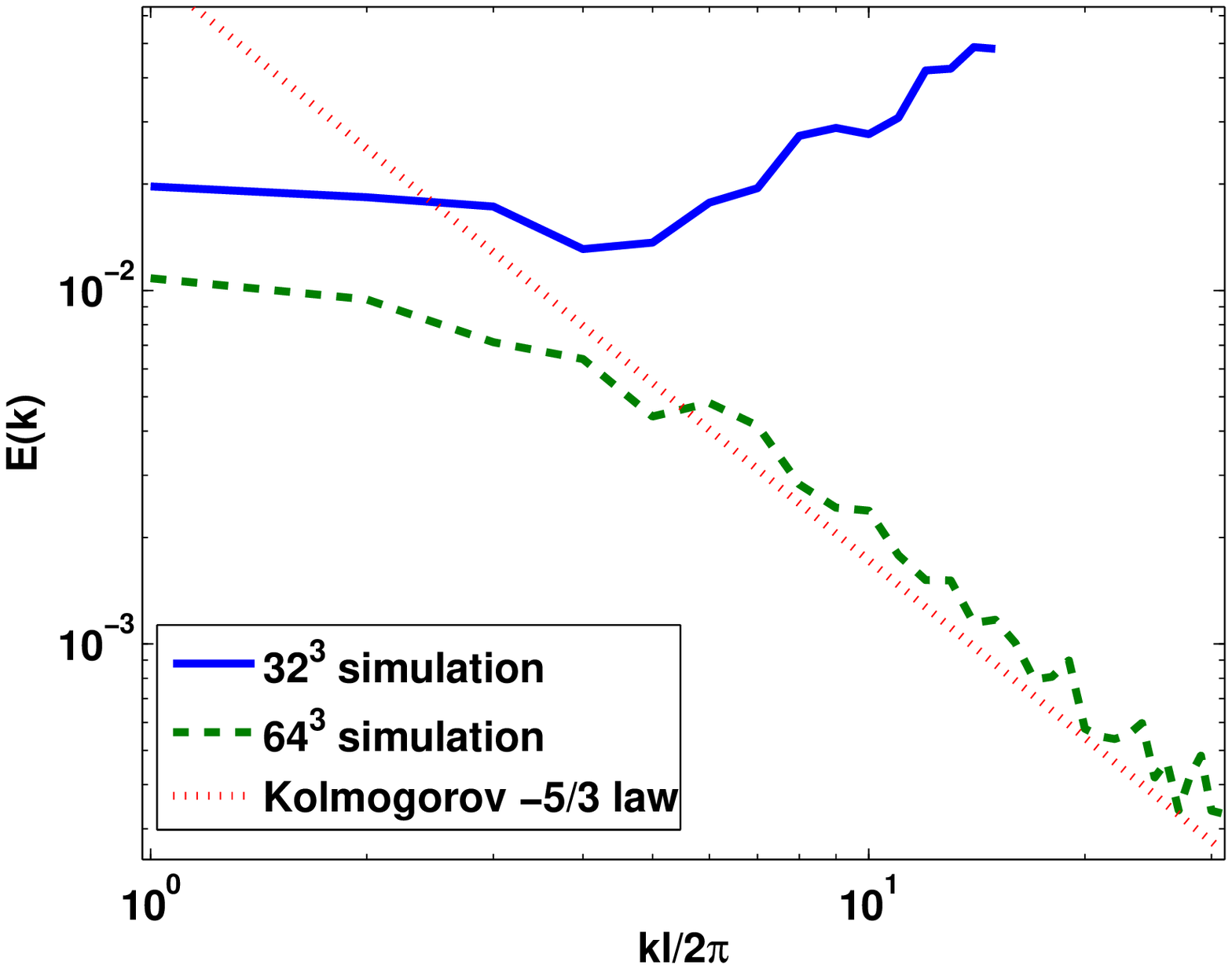}
   \caption{\small Energy spectra (of the velocity deviation from the
   laminar flow), for two different resolutions ($32^2$ and
   $64^3$). The rotation number is $R_\Omega=-1.016$ in all cases.
   Top panel: $Re=6000$. Middle panel: $Re=12000$. Bottom panel:
   $Re=20000$. At this resolution only the top panel simulations are
   resolved, as expected. See text for discussion.}{}\label{spectres}
\end{figure}

Comparing with Eq.~(\ref{ldef}), it appears that $\tau_d(k)$
represents the fraction of dissipation due to scales $|k_x|<k$,
$|k_y|<k$ and $|k_z|<k$. This quantity is plotted on
Fig.~\ref{spectre_dissip} with the $64^3$ simulation spectrum. It
appears that more than 95\% of the total dissipation is due to
$k<1/2 k_{\textrm {max}}$ (i.e., the $32^3$ resolution). Also,
comparing Fig.~\ref{spectre_dissip} with the top panel of
Fig.~\ref{spectres} indicates that most of the dissipation comes
from the part of the spectrum which is steeper than the Kolmogorov
spectrum, as one would expect.

It is also instructive to examine the spectral behavior at
Reynolds number larger than the transition Reynolds number, as
shown on Fig.\ref{spectres}.

This figure displays energy spectra of the velocity deviation from
the laminar flow. The rotation number is fixed at
$R_\Omega=-1.016$ for all spectra, and they have been averaged
over a 200 shear time period to reduce the noise. From top to
bottom, the Reynolds number is $6000$, $12000$ and $20000$
respectively. The $32^3$ simulations are resolved only in the top
panel, while the $64^3$ simulations should be resolved in the top
two panels. Comparing the second panel with the first reveals a
couple of interesting points:

\begin{itemize}
  \item The $64^3$ simulation shows an extension of the spectrum,
  compatible with a small inertial range (this is difficult to
  ascertain because of the remaining noise in the simulation),
  while still resolving at least the top of the dissipation range,
  but marginally so.
  \item The $32^3$ simulation begins to significantly deviate from
  the $64^3$ simulation, although the trend is similar.
\end{itemize}

The third panel also displays a fairly relevant piece of
information. The $64^3$ simulation shows both the self-sustaining
mechanism scales and the inertial spectra. However, the
dissipation scale does not seem to be resolved. This is not
unexpected, since increasing the Reynolds number necessarily
increases the inertial spectrum, and therefore decreases the
dissipation scale. Apparently, the dissipation scale is probably
not far off the resolved scales, so that the simulation
nevertheless does not noticeably deviate from the expected
behavior. But note that the $32^3$ simulation is clearly strongly
unresolved, with energy accumulating in the small scales in order
for a statistically steady equilibrium to be achieved: indeed, as this
simulation resolves the self-sustaining mechanism scale,
turbulence is present; however, as the smallest resolved scale is
significantly larger than the dissipation scale, the spectrum must be
strongly deformed to achieve a dissipation which is consistent
with the energy input due to the turbulence self-sustaining
mechanism.

These simulations illustrate that if the dissipation scale is not
resolved, the simulated flow does not necessarily relaminarize,
but the deformation of both the amplitude and shape of the
spectrum most likely results in, e.g., unreliable estimates of the
turbulent transport. In particular, the reliability of finite
difference simulations where no viscous term is explicitly
included in the code is unclear\footnote{We did not further
investigate this question here.}.

On the other hand, this point is related to the fact that the
numerical dissipation in a Fourier code is extremely weak, so that
the deformation of the spectrum may be quite large. To compute the
numerical dissipation explicitly, we have estimated its effect on
the turbulent energy budget.

\begin{figure}
   \centering
   \includegraphics[scale=0.30]{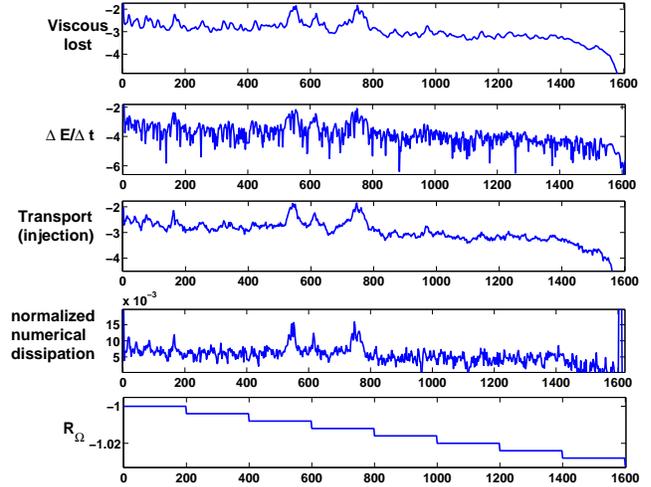}
   \caption{\small Energy budget for a $64^3$ $Re=20000$ run with our Fourier
   code. Each plot represent a term in Eq.~(\ref{w2avb}). The numerical
   dissipation is normalized by the total dissipation ${\partial}\langle
   {\bm w}^2/2\rangle/{\partial t} - S\langle w_x w_y\rangle$. We find that
   numerical dissipation is about 1\% of the total
   dissipation.}{}\label{energ-budg}
\end{figure}

We plot an example of such an energy budget on
Fig.~\ref{energ-budg}, where all the terms in Eq.~(\ref{w2av}) are
evaluated, so that the remaining difference measures the code
dissipation. One should note that these plots are integrated over
2 shear times, so that they include the numerical dissipation due
to the desaliazing procedure (done at each time loop) and losses
from the remapping procedure (done each shear time). The presented
result is generic: for all our simulations, numerical dissipation
is found to be at most a few percent of the total dissipation.

In summary, we have tried as much as possible to ensure that our
determination of the transition Reynolds number and turbulent
transport do not suffer from lack of resolution of the dissipation
scale. Note also that the results of the Fourier and finite
difference codes are consistent with each other. This makes us
confident that our simulations faithfully represent the relevant
physics, down to and including the dissipation scale, within the
relevant $(Re,R_\Omega)$ domain determined at each resolution on
Fig.~\ref{graphacyc}.

\subsubsection{Shearing sheet simulations and scale invariance}

Recently, \citet{B04} has argued that the scale invariance of the
inviscid form of the Navier-Stokes equation used here makes any
small scale solution exist at large scales as well. This argument
seems to imply that simulations of the kind performed here should
not be resolution limited. However, neither the simulations of
\cite{BHS96}, \cite{HBW99}, the ones performed here, nor a real
disk, are scale invariant. In shearing sheet simulations, the box
size defines a scale; in a real disk, the disk scale height does.
Furthermore, we point out that the mechanism analyzed by
\cite{W97}, whose qualitative and semi-quantitative relevance to
the present work has been pointed out hereabove, is somewhat
insensitive to the nature of the imposed boundary condition. Along
with the results found in this paper, this suggests that only a
scale rather than a specific boundary condition needs to be
imposed for statistically stationary turbulence to show up in
numerical simulations, as exemplified in section
\ref{res-anticyc}. Finally, the role of an increasingly dominant
Coriolis force is not to define another scale, which it cannot,
but to modify the relative range of scales that are required for
turbulence to exist (most likely because of its more and more
stringent time-scale requirement), so that numerical resolution
\emph{does} play an important role in subcritical turbulence
detection, as can be seen from Fig.~\ref{graphacyc}.

\subsection{Boundary conditions and aspect ratio}\label{BC}

Assessing the role of boundary conditions on the existence and
properties of subcritical turbulence is an important question,
since real accretion disk boundary conditions are not reproducible
in experimental flows. However, the resolution demand in the local
shearing box is already so large for a keplerian flow that a
global simulation of a subcritical keplerian disk flow is totally
out of reach. The best we can do is to compare numerical
experiments with shearing sheet and rigid/periodic boundary
conditions with one another, and with experimental results. This
is the object of this section.

Before doing so, let us point out some important differences
between the two types of boundary conditions:

\begin{itemize}
  \item In the semi-lagrangian variables ~defined by
  Eqs.~(\ref{shearaxes}), the only difference is that the
  velocity deviation from the laminar flow cancels on the rigid boundary
  in the rigid/periodic case, while it is periodic in the shear
  direction (as in the others) in the shearing sheet case. This
  results in a suppression of the boundary layer in the shearing
  sheet case.
  \item Characteristic sizes are the same in both cases. However,
  while for rigid/periodic conditions, structures are forced to
  remain more or less stationary with respect to the walls on average, this is
  not the case with shearing sheet boundary conditions, where structures can move
  at random through the boundary. As a
  consequence, a long-lasting mean flow distortion is apparent
  with rigid/periodic boundary conditions (due to the matching of
  turbulently enhanced transport with the viscous one in the boundary
  layer), while in shearing-sheet simulations, although such a distortion
  is usually locally found at any given time,
  it averages out over time, due to its random localization.
  \item This relates to a profound difference between accretion
  disks and actual experiments. In the latter, the flow profile
  adjusts to the imposed boundary condition through a pressure redistribution,
  and a stationary state is reached. In the former, this cannot
  take place, and the disk is never stationary, due to the
  resulting turbulent transport of mass and angular momentum.
\end{itemize}

In spite of these differences , we shall nevertheless argue that
the choice of boundary conditions has only a limited impact on
some of our qualitative and semi-quantitative results. This
suggests that the underlying mechanisms are reasonably closely
related in both settings, although much more work than what has
been possible to do here is required to ascertain this conclusion.

\subsubsection{Cyclonic rotation}\label{cycrot}

\begin{figure}
   \centering
   \includegraphics[scale=0.30]{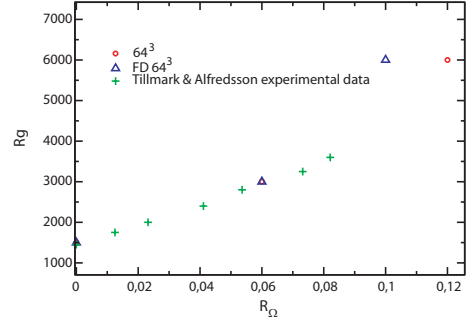}
   \caption{\small $Rg(R_\Omega)$ plot from experimental data (\citealt{TA96},
   crosses), and our numerical simulations using $64^3$ Fourier code (circles)
   and $64^3$ finite difference code (triangles) with cubic box and shearing
   sheet boundary conditions.}\label{TA-num}
\end{figure}

Fig.~\ref{TA-num} displays a comparison of our numerical results
with the \cite{TA96} data, in the range of rotation number where
these data were collected.

The agreement between the two is fair, with the Fourier code
results being sensibly more compatible with the data than the
finite difference code ones, at the larger rotation numbers. This
follows because, at the same ``resolution", a Fourier code is
physically more resolved than a Finite difference code. Note also
that some $128^3$ simulations were performed using the Fourier
code and the same transition thresholds were found as for the
$64^3$ simulations. This supports the idea that the $64^3$ Fourier
code results are not resolution limited.

\begin{figure}
  \centering
  \includegraphics[scale=0.29]{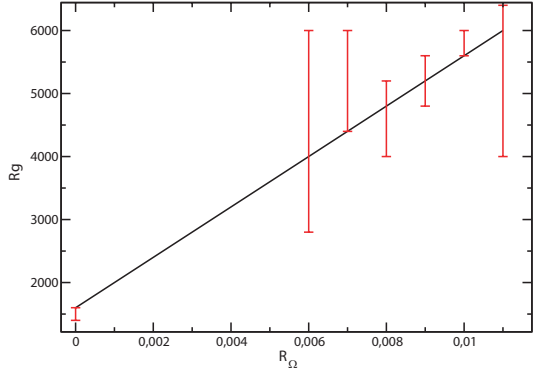}
  \caption{\small $Re_c$ as a function of $R_\Omega$ for
  cyclonic rotating plane couette flow.}\label{cyclocouette}
\end{figure}

We have also made a few runs using rigid (shearwise direction) and
periodic (other directions) boundary conditions with our ZEUS-like
code. At each rotation number, we made a few tries with different
Reynolds numbers to locate the transition threshold. Each run was
computed from the same initial condition for 400 shear times with
$80\times 80\times 40$ grid points and a $L_x=1.75\pi \, L_y=1 \,
L_z=1.2\pi$ aspect ratio box (corresponding to the ``minimal flow
unit" aspect ratio, i.e. the smallest box in which turbulence can
be sustained with these boundary conditions: see \citealt{HKW95}
for details). The error bars upper bounds correspond to the lower
Reynolds for which turbulence is found and the lower bound the
higher Reynolds number for which turbulence is lost. The numerical
data are shown on Fig.~\ref{cyclocouette}; the error bars reflect
our poor sampling, not intrinsic fluctuations in the transition
Reynolds number. These data are fitted by a linear law:

\begin{equation}
Rg=1400+4\times 10^5 R_\Omega,
\end{equation}

\noindent the slope of which is 15 times steeper than the one
found from the experimental data.

This dramatic difference in transition Reynolds number with
respect to the experimental and shearing sheet results is in fact
controlled by the choice of the simulation box aspect ratio. For
example, let us choose a longer box in the $z$ direction (i.e
$L_x=1.75\pi \, L_y=1 \, L_z=2.4\pi$). With such a choice,
turbulence is sustained at $R_\Omega=0.01$ and $Re=2400$, much
closer to the expected transition Reynolds of Fig.~\ref{TA-num}
than what is predicted by Fig.~\ref{cyclocouette}. Finally,
\cite{KLJ96}, using a very elongated simulation box in the flow
direction, found transition right at the experimentaly determined
Reynolds number ($Rg=3000$, $R_\Omega=0.06$).

These result show the important role of aspect ratio in
subcritical turbulence simulations with rigid/periodic boundary
conditions. Apparently, the use of shearing sheet boundary
conditions relaxes this constraint. This is reasonable since the
shearing sheet box allows more freedom than rigid boundary
conditions. In actual experiments, the aspect ratio is not an
issue since usually very large $L_x/L_y$ and $L_z/l_y$ are used,
so that the turbulence coherence length can freely adjust itself
in these directions.

These results also indirectly suggest that the turbulence
saturation mechanism is not strongly affected by the use of
shearing sheet boundary conditions. One would nevertheless expect
that the reduction of the shear in the middle of the flow, due to
the mean velocity profile modification which occurs with
rigid/periodic boundary conditions, produces a reduced turbulent
transport. This is indeed the case: e.g., the turbulent transport
at marginal stability ($R_\Omega=0$) is $\langle v_x
v_y\rangle\simeq 2\times 10^{-3} (Sd)^2$ for the rigid/periodic
boundary conditions\footnote{This is measured in the middle of the
flow where the turbulent transport is maximized, and viscous
transport negligible.}, while one has $\langle v_x
v_y\rangle\simeq 0.4 (Sd)^2$ throughout the flow with the shearing
sheet boundary conditions, although the transition Reynolds number
is the same in both instances. These features most probably find a
natural explanation if the turbulence amplitude saturation
mechanism is primarily controlled by the system nonlinearities,
and not by the mean profile deformation.

\subsubsection{Anticyclonic rotation}

\begin{figure}
  \centering
  \includegraphics[scale=0.4]{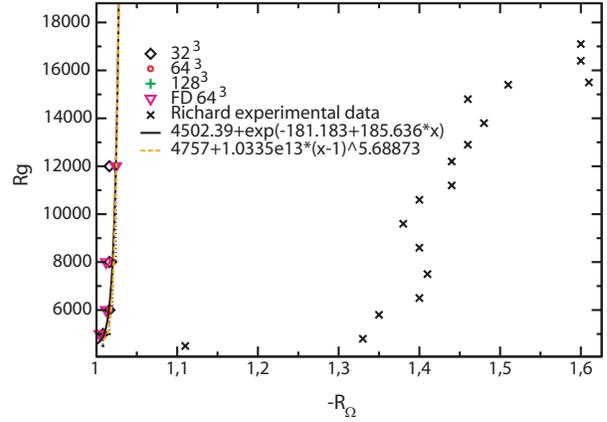}
  \caption{\small $Rg(R_\Omega)$ plot from experimental data on Taylor-Couette
  flows (\citealt{R01}, crosses), and the various numerical simulations results
  and related fits shown on Fig.~\ref{graphacyc}.}\label{anticyc2}
\end{figure}

The comparison of our numerical results with \cite{R01} data is
shown on Fig.~\ref{anticyc2}.

The discrepancy between the experimental and numerical data is
striking, especially at the light of the remarkable consistency
observed for cyclonically rotating flows. In particular, the
increase in transition Reynolds is considerably steeper in the
numerical data than in the experimental ones. Note however that
the numerical and experimental data seem to give the same
transition Reynolds number \textit{at the marginal stability
boundary}.

\smallskip

\cite{LD05} have argued that the flow curvature plays no role in
the anticyclonic flow data of \cite{R01}, so that the origin of
the large discrepancy between the numerical and experimental
results must be found elsewhere\footnote{In any case, the flow
curvature always increases the transition Reynolds number, so that
including curvature in the analysis of this problem can only make
it worse, not better.}. In this respect, note that experimental
secondary flow distortions are much more likely to induce a linear
instability somewhere in the flow on the anticyclonic side as on
the cyclonic one. Indeed, recall that the stability limit is
defined by Eq.~(\ref{stab}). Consider the cyclonic marginal
stability limit first ($R_\Omega=0$), and assume that one moves
away from it by imposing a small change in rotation
$\delta\Omega$. The required change in shear profile $\delta S$ to
locally achieve $2\delta\Omega/(S(y)+\delta S) < 0$ is large:
$\delta S\sim S(y)$ is needed. Conversely, at the anticyclonic
marginal stability limit ($R_\Omega=-1$, i.e., $S=2\Omega$), upon
a small change $\delta\Omega$ of the rotation rate, a change
$\delta S\simeq2\delta\Omega \ll S$ suffices to locally make
$2\Omega/S > -1$ and produce a linear instability somewhere in the
system. This argument shows that the presence of secondary flows,
such as Ekmann's circulation, can easily make the flow more
unstable than it would be in its absence in anticyclonic flows,
whereas this is much more difficult to achieve in cyclonic ones.
This may easily explain the discrepancy between numerical and
experimental results shown on Fig.~\ref{anticyc2}, while the
agreement is remarkable at the marginal stability boundary.

\section{Summary and conclusions}\label{conclusion}

The central results of this paper are displayed on
Figs.~\ref{graphcyc}, \ref{transcyc}, \ref{graphacyc},
\ref{transacyc} and \ref{spectrum}, and their significance and
implications are discussed in sections \ref{res-cyc},
\ref{res-anticyc}, \ref{phen}, \ref{kep}, \ref{cor}, and
\ref{eff-re}. The main implications of these results are
summarized in the abstract. In the course of the discussion, we
have found that a number incorrect statements have been made in
the literature, most notably concerning the existence and
importance of subcritical turbulence in presence of a dynamically
significant Coriolis force. We have also found that resolution is
a key issue for subcritical anticyclonically rotating flows
(including keplerian ones), and have quantified the relation
between resolution, rotation and Reynolds number. In relation to
this, we believe that the question of resolution of the
dissipation scale is not emphasized enough in the astrophysics
literature, and the potential effects of this problem are most
probably underestimated.

Our simulations do not faithfully represent a real disk:
neither vertical stratification, nor, more critically, realistic
vertical boundary conditions have been implemented. A real
(hydrodynamic disk) moves either in vacuum, or, more probably, in
a hot corona. In both cases, one expects the real vertical
boundary condition in the disk to be (nearly) stress-free. We have
made some very preliminary simulations of ah stratified
disk-corona system to test this idea, where most of the inertia
lies in the disk. Although a strong numerical mixing of the corona
and the disk at the interface prevents us to evolve the system for
a long time ($50 t_s$ max), no significant difference in the
overall dynamics of the disk did show up. However this problem
probably requires more careful investigations to validate this
conclusion.

Overall, the outcome of this investigation still leaves us with
the issue of transport unresolved in MHD-inactive flows (and
possibly in some MHD-active ones), and we will briefly comment the
various ways out of this conundrum.

We first note that an efficient enough local instability should
lead to a large enough turbulent transport, because the transition
to fully developed turbulence usually occurs to significantly
lower Reynolds numbers in these systems than the ones found here.
This is true, e.g., in rotating shear flows of the type considered
here, in the linearly unstable regime. However, no such
instability has yet been found in hydrodynamical keplerian flows,
either stratified or not, as discussed in the introduction.
It remains to be seen whether another such instability can
operate in hydrodynamic disks, but the list of potential driving
agents has by now significantly been narrowed.

In what concerns the YSO disks dead-zone in particular, it may be
that the disk stirring due to the MRI above and below the
dead-zone itself \citep{FS03} might provide enough transport in
the end if it excites large enough large scale 2D disturbances of
the right type \citep{IK01} in the disk. However, this option
remains to be worked out in detail.

It has often been noted that transport in disks may not be due to
turbulence but to waves (see, e.g., \citealt{PL95} for an
introduction to the subject). Recent results on the existence of
vortices in stratified disks \citep{BM05} and on the coupling of
waves to vortices resulting in efficient transport in 2D dynamics
(\citealt{BCMTRF05} and references therein) support this idea.

%However, this option is excluded in disk driving jets. In such a
%context, most of the transport is usually vertical (in the jet)
%rather than radial (in the disk). Such configurations are often
%criticized on two different fronts. First, it has been pointed out
%that in a standard disk model supplied with a little diffusivity,
%the field line are either not bent enough \citep{LPP94a}, or the
%flux is expelled \citep{HPB96}. However, global disk-wind
%solutions do exist, in which the inflow is substantially larger
%than the standard model prediction \citep{L96,F97}, and with
%sufficient resultant field line bending. Secondly, disk-jet models
%have been argued to be unstable \citep{LPP94b,CS02}; however, the
%global disk-wind solutions just mentioned may escape this
%argument, as they produce a less important wind for more highly
%inclined field lines due to the role of the vertical magneric
%field pressure. Turbulence is critically needed in these
%structures, as these global solutions rely on a rather high
%turbulent resistivity for their existence. They also require a
%quasi-equipartition near the disk mid-plane ($\beta\sim 1$), and
%it is unclear in such a context whether the magneto-rotational
%instability (MRI) is not quenched. If this is the case, another
%(M)HD instability must also be looked for as a source of local
%turbulence in these structures. In any case, the overall outcome
%and efficiency of the MRI is still not well-known \citep{MS00} in
%this configuration, and requires further study.

\appendix

\section{Displaced particle analysis for rotating
flows:}\label{app:disp}

The following line of argument closely follows \citet{TD81} and
\citet{T92}. Let us consider a rotating shear flow, whose dynamics
is controlled by Eq.~(\ref{split-RPC}). As in section \ref{equ},
$x$ is the direction of the flow, $y$ the direction of the shear,
and $z$ the direction perpendicular to the $x,y$ plane, in which
the rotation $\bf\Omega$ is applied. The laminar equilibrium
velocity ${\bf u}_L= (U(y), 0, 0)$ generates a Coriolis force in
the $y$ direction of magnitude $-2\rho\Omega U$ (in algebraic
value), which is balanced by the equilibrium generalized pressure
gradient $-d \pi/dy$.

\begin{figure}[htb]
  \centering
  \includegraphics[scale=0.7]{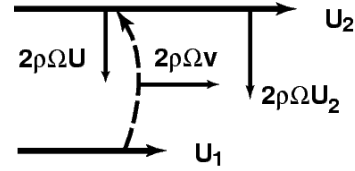}
  \caption{\small Sketch of the effect of the Coriolis force on the displaced particle
  (see text).}\label{graphe}
\end{figure}

Let us further consider two fluid ``rods" of infinite extent in
the streamwise direction $x$, and located at positions $y_1$ and
$y_2=y_1+\delta y$. The streamwise velocities of these rods are
$U_1$ and $U_2$, respectively. Let us assume that one displaces
the rod at $y_1$ to location $y_2$, without disturbing the
pressure distribution. Although the total work of the Coriolis
force vanishes, there is a net partial work due to the force
component in the $x$ direction which originates in the velocity
$v$ of this displacement in the $y$ direction. Because of this
partial work, the rod experiences a change of $x$ momentum, and
therefore of $x$ velocity, which reads

\begin{equation}\label{Corwork}
U_1'-U_1=\int 2\Omega v dt=2\Omega\delta y,
\end{equation}

\noindent so that the velocity $U_1'$ of the rod when it reaches
location $y_2$ differs from the equilibrium velocity $U_2$, and
correlatively, the $x$ component of the Coriolis force acting on
this displaced rod, $-2\rho\Omega U_1'$ (in algebraic value)
differs from the equilibrium one, $-2\rho\Omega U_2$ (see
Fig.~\ref{graphe}).

Consequently, the net result between the equilibrium pressure
gradient and the Coriolis force will tend to restore the displaced
rod to its equilibrium position\footnote{Consider the direction
and magnitude of the Coriolis force and pressure force along the
$y$ axis to derive the effect of the inequality.} if $U_1'> U_2$,
or displace it further if $U_1'<U_2$. From Eqs.~(\ref{Corwork})
and (\ref{rot}), one obtains

\begin{equation}\label{difvel}
U_1'-U_2=2\Omega\delta y - \frac{dU}{dy}\delta
y=S(R_\Omega+1)\delta y.
\end{equation}

\noindent where $S=-dU/dy$ is the shear. From this result, the net
force (Coriolis and pressure) on the displaced rod reads

\begin{equation}\label{netforce}
  2\rho\Omega(U_2-U_1')=-\rho S^2
  R_\Omega(R_\Omega+1)\delta y,
\end{equation}

\noindent This shows that equilibrium is always restored when
$R_\Omega>0$ or $R_\Omega<-1$ and destroyed otherwise (equality
holds at marginal stability). This is the result quoted in section
\ref{equ}. This result can also be directly derived from the
linearized eulerian equation of motion with the use of spatially
uniform perturbations of the pressure and the velocity.

\begin{acknowledgements}

The undertaking and completion of this work has benefitted from
discussions with a number of colleagues in the past few years,
most notably O.\ Dauchot, F.\ Daviaud, B.\ Dubrulle, F.\
Ligni\`eres, and J.-P. Zahn. PYL also acknowledges discussions
held at the KITP conference on the physics of astrophysical
outflows and accretion disks, with C.\ Gammie, J.\ Hawley, B.\
Johnson, E.\ Quataert, and J.\ Goodman. Interactions with O.\
Blaes and G.\ Bodo have been particularly fruitful. David Clarke
has provided us with his version of the ZEUS code, and his
friendly help (as well as G.\ Bodo's) in the initial phases of
this project is gratefully acknowledged. We thank Fran\c coise
Roch, Fran\c coise Berthoud and Alain Pasturel for their help in
accessing the computing resources of the SCCI of the Grenoble
Observatory, and of the PHYNUM and MEDETPHY platforms of the
Grenoble University CIMENT project. A large fraction of the
simulations presented here has also been performed at IDRIS
(French national computational center).

\end{acknowledgements}

\bibliographystyle{aa}

\bibliography{3683gles}

\begin{thebibliography}{85}
\expandafter\ifx\csname natexlab\endcsname\relax\def\natexlab#1{#1}\fi

\bibitem[{{Afshordi} {et~al.}(2004){Afshordi}, {Mukhopadhyay}, \&
  {Narayan}}]{AMN05}
{Afshordi}, N., {Mukhopadhyay}, B., \& {Narayan}, R. 2004, \textit{Accepted in}
  \apj, astro-ph/0412194

\bibitem[{{Arlt} \& {Urpin}(2004)}]{AU04}
{Arlt}, R. \& {Urpin}, V. 2004, \aap, 426, 755

\bibitem[{{Balbus}(2003)}]{B03}
{Balbus}, S.~A. 2003, \araa, 41, 555

\bibitem[{{Balbus}(2004)}]{B04}
---. 2004, \textit{Submitted to} \aap, astro-ph/0408510

\bibitem[{{Balbus} \& {Hawley}(1991)}]{BH91}
{Balbus}, S.~A. \& {Hawley}, J.~F. 1991, \apj, 376, 214

\bibitem[{{Balbus} {et~al.}(1996){Balbus}, {Hawley}, \& {Stone}}]{BHS96}
{Balbus}, S.~A., {Hawley}, J.~F., \& {Stone}, J.~M. 1996, \apj, 467, 76

\bibitem[{{Barranco} \& {Marcus}(2005)}]{BM05}
{Barranco}, J.~A. \& {Marcus}, P.~S. 2005, \textit{Submitted to} \apj,
  astro-ph/0501267

\bibitem[{{Bech} \& {Andersson}(1997)}]{BA97}
{Bech}, K.~H. \& {Andersson}, H.~I. 1997, J.\ Fluid Mech., 347, 289

\bibitem[{{Blaes}(1987)}]{B87}
{Blaes}, O.~M. 1987, \mnras, 227, 975

\bibitem[{{Bodo} {et~al.}(2005){Bodo}, {Chagelishvili}, {Murante}, {Tevzadze},
  {Rossi}, \& {Ferrari}}]{BCMTRF05}
{Bodo}, G., {Chagelishvili}, G., {Murante}, G., {et~al.} 2005, \textit{Accepted
  in} \aap, astro-ph/0503474

\bibitem[{{Bottin} {et~al.}(1997){Bottin}, {Dauchot}, \& {Daviaud}}]{BDD97}
{Bottin}, S., {Dauchot}, O., \& {Daviaud}, F. 1997, Phys.\ Rev.\ Lett., 79,
  4377

\bibitem[{{Brandenburg} \& {Dintrans}(2001)}]{BD01}
{Brandenburg}, A. \& {Dintrans}, B. 2001, astro-ph/0111313

\bibitem[{{Brosa} \& {Grossmann}(1999)}]{BG99}
{Brosa}, U. \& {Grossmann}, S. 1999, European Phys.\ J.\ B, 9, 343

\bibitem[{{Cabot}(1996)}]{C96}
{Cabot}, W. 1996, \apj, 465, 874

\bibitem[{{Cambon} {et~al.}(1994){Cambon}, {Benoit}, {Shao}, \&
  {Jacquin}}]{CBSJ94}
{Cambon}, C., {Benoit}, J., {Shao}, L., \& {Jacquin}, L. 1994, J.\ Fluid Mech.,
  278, 175

\bibitem[{{Chagelishvili} {et~al.}(2003){Chagelishvili}, {Zahn}, {Tevzadze}, \&
  {Lominadze}}]{CZTL03}
{Chagelishvili}, G.~D., {Zahn}, J.-P., {Tevzadze}, A.~G., \& {Lominadze}, J.~G.
  2003, \aap, 402, 401

\bibitem[{{Darbyshire} \& {Mullin}(1995)}]{DM95}
{Darbyshire}, A.~G. \& {Mullin}, T. 1995, J.\ Fluid Mech., 289, 83

\bibitem[{{Dauchot} \& {Daviaud}(1995{\natexlab{a}})}]{DD95a}
{Dauchot}, O. \& {Daviaud}, F. 1995{\natexlab{a}}, Phys.\ Fluids, 7, 335

\bibitem[{{Dauchot} \& {Daviaud}(1995{\natexlab{b}})}]{DD95b}
---. 1995{\natexlab{b}}, Phys.\ Fluids, 7, 901

\bibitem[{{Daviaud} {et~al.}(1992){Daviaud}, {Hegseth}, \& {Berg\'e}}]{DHB92}
{Daviaud}, F., {Hegseth}, J., \& {Berg\'e}, P. 1992, Phys.\ Rev.\ Lett., 69,
  2511

\bibitem[{{Drazin} \& {Reid}(1981)}]{DR81}
{Drazin}, P. \& {Reid}, W. 1981, Hydrodynamic stability (Cambridge Univ.\
  Press)

\bibitem[{{Dubrulle}(1993)}]{D93}
{Dubrulle}, B. 1993, Icarus, 106, 59

\bibitem[{{Dubrulle} {et~al.}(2005{\natexlab{a}}){Dubrulle}, {Dauchot},
  {Daviaud}, {Longaretti}, {Richard}, \& {Zahn}}]{DDDLRZ05}
{Dubrulle}, B., {Dauchot}, O., {Daviaud}, F., {et~al.} 2005{\natexlab{a}},
  \textit{Accepted} in Phys.\ Fluids

\bibitem[{{Dubrulle} {et~al.}(2005{\natexlab{b}}){Dubrulle}, {Mari{\' e}},
  {Normand}, {Richard}, {Hersant}, \& {Zahn}}]{DMNRHZ05}
{Dubrulle}, B., {Mari{\' e}}, L., {Normand}, C., {et~al.} 2005{\natexlab{b}},
  \aap, 429, 1

\bibitem[{{Dubrulle} \& {Zahn}(1991)}]{DZ91}
{Dubrulle}, B. \& {Zahn}, J.-P. 1991, Journal of Fluid Mechanics, 231, 561

\bibitem[{{Eckhardt} \& {Mersmann}(1999)}]{EM99}
{Eckhardt}, B. \& {Mersmann}, A. 1999, \pre, 60, 509

\bibitem[{{Faisst} \& {Eckhardt}(2004)}]{FE05}
{Faisst}, H. \& {Eckhardt}, B. 2004, J.\ Fluid Mech., 504, 343

\bibitem[{{Fleming} \& {Stone}(2003)}]{FS03}
{Fleming}, T. \& {Stone}, J.~M. 2003, \apj, 585, 908

\bibitem[{{Gammie}(1996)}]{G96}
{Gammie}, C.~F. 1996, \apj, 457, 355

\bibitem[{{Garaud} \& {Ogilvie}(2005)}]{GO05}
{Garaud}, P. \& {Ogilvie}, G.~I. 2005, \textit{Submitted to} J.\ Fluid Mech.,
  astro-ph/0503223

\bibitem[{{Goldreich} \& {Schubert}(1967)}]{GS67}
{Goldreich}, P. \& {Schubert}, G. 1967, \apj, 150, 571

\bibitem[{{Goodman} \& {Balbus}(2001)}]{GB01}
{Goodman}, J. \& {Balbus}, S.~A. 2001, astro-ph/0110229

\bibitem[{{Grossman}(2000)}]{G00}
{Grossman}, S. 2000, Rev.\ Mod.\ Phys., 72, 603

\bibitem[{{Hamilton} {et~al.}(1995){Hamilton}, {Kim}, \& {Waleffe}}]{HKW95}
{Hamilton}, J.~M., {Kim}, J., \& {Waleffe}, F. 1995, J.\ Fluid Mech., 287, 317

\bibitem[{{Hawley}(1991)}]{H91}
{Hawley}, J.~F. 1991, \apj, 381, 496

\bibitem[{{Hawley} {et~al.}(1999){Hawley}, {Balbus}, \& {Winters}}]{HBW99}
{Hawley}, J.~F., {Balbus}, S.~A., \& {Winters}, W.~F. 1999, \apj, 518, 394

\bibitem[{{Hawley} {et~al.}(1995){Hawley}, {Gammie}, \& {Balbus}}]{HGB95}
{Hawley}, J.~F., {Gammie}, C.~F., \& {Balbus}, S.~A. 1995, \apj, 440, 742

\bibitem[{{Hersant} {et~al.}(2005){Hersant}, {Dubrulle}, \& {Hur{\'
  e}}}]{HDH05}
{Hersant}, F., {Dubrulle}, B., \& {Hur{\' e}}, J.-M. 2005, \aap, 429, 531

\bibitem[{{Ioannou} \& {Kakouris}(2001)}]{IK01}
{Ioannou}, P.~J. \& {Kakouris}, A. 2001, \apj, 550, 931

\bibitem[{{Johnson} \& {Gammie}(2005{\natexlab{a}})}]{JG05a}
{Johnson}, B.~M. \& {Gammie}, C.~F. 2005{\natexlab{a}}, Submitted to \apj,
  astro-ph/0501005

\bibitem[{{Johnson} \& {Gammie}(2005{\natexlab{b}})}]{JG05b}
---. 2005{\natexlab{b}}, In preparation

\bibitem[{{Klahr}(2004)}]{K04}
{Klahr}, H. 2004, \apj, 606, 1070

\bibitem[{{Klahr} \& {Bodenheimer}(2003)}]{KB03}
{Klahr}, H.~H. \& {Bodenheimer}, P. 2003, \apj, 582, 869

\bibitem[{{Komminaho} {et~al.}(1996){Komminaho}, {Lundbladh}, \&
  {Johansson}}]{KLJ96}
{Komminaho}, J., {Lundbladh}, A., \& {Johansson}, A.~V. 1996, J.\ Fluid Mech.,
  320, 259

\bibitem[{{Leblanc} \& {Cambon}(1997)}]{LC97}
{Leblanc}, S. \& {Cambon}, C. 1997, Phys.\ Fluids, 9, 1307

\bibitem[{{Lerner} \& {Knobloch}(1988)}]{LK88}
{Lerner}, J. \& {Knobloch}, E. 1988, Journal of Fluid Mechanics, 189, 117

\bibitem[{{Lesieur}(1990)}]{L90}
{Lesieur}, M. 1990, Turbulence in fluids third edition (Kluwer)

\bibitem[{{Longaretti}(2002)}]{L02}
{Longaretti}, P.-Y. 2002, \apj, 576, 587

\bibitem[{{Longaretti} \& {Dauchot}(2005)}]{LD05}
{Longaretti}, P.-Y. \& {Dauchot}, O. 2005, in {Proc.\ of the Bristol 2004
  Symposium on the laminar-turbulent transition}, ed. {Kerswell and Mullin}
  ({Kluwer})

\bibitem[{{Mukhopadhyay} {et~al.}(2004){Mukhopadhyay}, {Afshordi}, \&
  {Narayan}}]{MAN05}
{Mukhopadhyay}, B., {Afshordi}, N., \& {Narayan}, R. 2004, \textit{Submitted
  to} \apj, astro-ph/0412193

\bibitem[{{Ogilvie}(2003)}]{O03}
{Ogilvie}, G.~I. 2003, \mnras, 340, 969

\bibitem[{{Papaloizou} \& {Lin}(1995)}]{PL95}
{Papaloizou}, J.~C.~B. \& {Lin}, D.~N.~C. 1995, \araa, 33, 505

\bibitem[{{Papaloizou} \& {Pringle}(1984)}]{PP84}
{Papaloizou}, J.~C.~B. \& {Pringle}, J.~E. 1984, \mnras, 208, 721

\bibitem[{{Pedley}(1969)}]{P69}
{Pedley}, T.~J. 1969, J.\ Fluid Mech., 35, 97

\bibitem[{{Peyret}(2002)}]{P02}
{Peyret}, R. 2002, Spectral Methods for Incompressible Viscous Flow (Springer)

\bibitem[{{Pumir}(1996)}]{P96}
{Pumir}, A. 1996, Phys.\ Fluids, 8, 3112

\bibitem[{{R{\" u}diger} {et~al.}(2002){R{\" u}diger}, {Arlt}, \&
  {Shalybkov}}]{RAS02}
{R{\" u}diger}, G., {Arlt}, R., \& {Shalybkov}, D. 2002, \aap, 391, 781

\bibitem[{{Richard}(2001)}]{R01}
{Richard}, D. 2001, PhD thesis, Universit\'e de Paris VII

\bibitem[{{Richard} {et~al.}(2001){Richard}, {Dauchot}, \& {Zahn}}]{RDDZ01}
{Richard}, D., {Dauchot}, O., \& {Zahn}, J.-P. 2001, in {Proc.\ of the 12th
  Couette-Taylor Workshop, Evanston, USA}

\bibitem[{{Richard} \& {Zahn}(1999)}]{RZ99}
{Richard}, D. \& {Zahn}, J. 1999, \aap, 347, 734

\bibitem[{{Rogallo}(1981)}]{R81}
{Rogallo}, R.~S. 1981, NASA STI/Recon Technical Report N, 81, 31508

\bibitem[{{Romanov}(1973)}]{R73}
{Romanov}, V.~A. 1973, Func.\ Anal.\ Appl., 7, 137

\bibitem[{{Salhi} \& {Cambon}(1997)}]{SC97}
{Salhi}, A. \& {Cambon}, C. 1997, J.\ Fluid Mech., 347, 171

\bibitem[{{Satomura}(1981)}]{S81}
{Satomura}, T. 1981, J.\ Meteor.\ Soc.\ Japan, 59, 148

\bibitem[{{Schmiegel} \& {Eckhardt}(1997)}]{SE97}
{Schmiegel}, A. \& {Eckhardt}, B. 1997, Phys.\ Rev.\ Lett., 79, 5250

\bibitem[{{Shakura} {et~al.}(1978){Shakura}, {Sunyaev}, \&
  {Zilitinkevich}}]{SSL78}
{Shakura}, N.~I., {Sunyaev}, R.~A., \& {Zilitinkevich}, S.~S. 1978, \aap, 62,
  179

\bibitem[{{Shalybkov} \& {Ruediger}(2005)}]{SR05}
{Shalybkov}, D. \& {Ruediger}, G. 2005, \aap, 438, 411

\bibitem[{{Sipp} \& {Jacquin}(2000)}]{SJ00}
{Sipp}, D. \& {Jacquin}, L. 2000, Phys.\ Fluids, 12, 1740

\bibitem[{{Speziale}(1991)}]{S91}
{Speziale}, C.~G. 1991, Ann.\ Rev.\ Fluid Mech., 23, 107

\bibitem[{{Speziale} \& {Mhuiris}(1989)}]{SMM89}
{Speziale}, C.~G. \& {Mhuiris}, N.~M. 1989, Phys.\ Fluids, 1, 294

\bibitem[{{Stone} \& {Balbus}(1996)}]{SB96}
{Stone}, J.~M. \& {Balbus}, S.~A. 1996, \apj, 464, 364

\bibitem[{{Stone} \& {Norman}(1992)}]{SN92}
{Stone}, J.~M. \& {Norman}, M.~L. 1992, \apjs, 80, 753

\bibitem[{{Taylor}(1936)}]{T36}
{Taylor}, G.~I. 1936, Proc.\ Roy.\ Soc.\ London A, 157, 546

\bibitem[{{Tevzadze} {et~al.}(2003){Tevzadze}, {Chagelishvili}, {Zahn},
  {Chanishvili}, \& {Lominadze}}]{TCZCL03}
{Tevzadze}, A.~G., {Chagelishvili}, G.~D., {Zahn}, J.-P., {Chanishvili}, R.~G.,
  \& {Lominadze}, J.~G. 2003, \aap, 407, 779

\bibitem[{{Tillmark} \& {Alfredsson}(1996)}]{TA96}
{Tillmark}, N. \& {Alfredsson}, P.~H. 1996, in {Advances in Turbulence VI.},
  ed. S.~{Gavrilakis}, L.~{Machiels}, \& P.~A. {Monkewitz} (Kluwer), 391--394

\bibitem[{{Tritton}(1992)}]{T92}
{Tritton}, D.~J. 1992, J.\ Fluid Mech., 241, 503

\bibitem[{{Tritton} \& {Davies}(1981)}]{TD81}
{Tritton}, D.~J. \& {Davies}, P.~A. 1981, in Hydrodynamic instabilities and the
  transition to turbulence (Springer-Verlag), 229--270

\bibitem[{{Umurhan}(2005)}]{U05}
{Umurhan}, O.~M. 2005, Submitted to \mnras, astro-ph/0506016

\bibitem[{{Umurhan} \& {Regev}(2004)}]{UR05}
{Umurhan}, O.~M. \& {Regev}, O. 2004, \aap, 427, 855

\bibitem[{{Urpin}(2003)}]{U03}
{Urpin}, V. 2003, \aap, 404, 397

\bibitem[{{Waleffe}(1995)}]{W95}
{Waleffe}, F. 1995, Phys.\ Fluids, 7, 3060

\bibitem[{{Waleffe}(1997)}]{W97}
---. 1997, Phys.\ Fluids, 9, 883

\bibitem[{{Waleffe}(2003)}]{W03}
---. 2003, Phys.\ Fluids, 15, 1517

\bibitem[{{Wendt}(1933)}]{W33}
{Wendt}, G. 1933, Ing.\ Arch., 4, 577

\bibitem[{{Yecko}(2004)}]{Y04}
{Yecko}, P.~A. 2004, \aap, 425, 385

\end{thebibliography}

\end{document}